\documentstyle[sprocl,epsf]{article}

\input{psfig.tex}
\def\NPB{{\em Nucl. Phys.} B}
\def\PLB{{\em Phys. Lett.}  B}
\def\PRL{{\em Phys. Rev. Lett.}}
\def\PRD{{\em Phys. Rev.} D}

\def\LNC{{\em Lett. Nuovo Cimento}}

\def\be{\begin{equation}}
\def\ee{\end{equation}}
\def\bea{\begin{eqnarray}}
\def\eea{\end{eqnarray}}

\newcommand{\gae}{$\stackrel{>}{\sim}$}
\newcommand{\beq}{\begin{equation}}
\newcommand{\eeq}{\end{equation}}
\newcommand{\beqa}{\begin{eqnarray}}
\newcommand{\eeqa}{\end{eqnarray}}

\def\slashchar#1{\setbox0=\hbox{$#1$}           
   \dimen0=\wd0                                 
   \setbox1=\hbox{/} \dimen1=\wd1               
   \ifdim\dimen0>\dimen1                        
      \rlap{\hbox to \dimen0{\hfil/\hfil}}      
      #1                                        
   \else                                        
      \rlap{\hbox to \dimen1{\hfil$#1$\hfil}}   
      /                                         
   \fi}                                         %

\newcommand{\tr}{{\rm Tr}}
\newcommand{\im}{{\rm i}}

\newcommand{\mev}{{\rm \, MeV}}
\newcommand{\gev}{{\rm \, GeV}}
\newcommand{\tev}{{\rm \, TeV}}
\newcommand{\W}{{\rm\bf W}}
\newcommand{\B}{{\rm\bf B}}

\newcommand{\pr}{Phys.\ Rev.\ }
\newcommand{\prp}{Phys.\ Rep.\ }

\newcommand{\pl}{Phys.\ Lett.\ {\bf B}}

\begin{document}

\title{An Introduction to Dynamical\\
Electroweak Symmetry Breaking}

\author{R. Sekhar Chivukula}

\address{Physics Department, Boston University\\
590 Commonwealth Ave., Boston MA 02215 USA\\
{\tt sekhar@bu.edu}\\
{\tt http://physics.bu.edu/$\sim$sekhar}\\
{\tt BUHEP-97-2 and hep-ph/9701322}}


\maketitle\abstracts{In these lectures, I present an introduction to 
the theory and phenomenology of dynamical electroweak symmetry breaking.}

\section{Lecture 1: The Dynamics of Electroweak Symmetry Breaking}

\subsection{What's Wrong with the Standard Model?}

In the standard higgs model, one introduces
a fundamental scalar doublet:
\beq
\phi=\left(\matrix{\phi^+ \cr \phi^0 \cr}\right)
{}~,
\eeq
with potential:
\beq
V(\phi)=\lambda \left(\phi^{\dagger}\phi - {v^2\over 2}\right)^2
{}~.
\label{eq:pot}
\eeq
While this theory is simple and renormalizable, it has
a number of shortcomings. First, while the theory
can be constructed to accommodate the
breaking of electroweak symmetry, it provides no {\it explanation} 
for it -- one simply assumes that the potential
is of the form in eqn. \ref{eq:pot}. In addition, in
the absence of supersymmetry, quantum corrections to
the Higgs mass are naturally of order the largest scale
in the theory
\beq
{\lower5pt\hbox{\epsfysize=0.25 truein \epsfbox{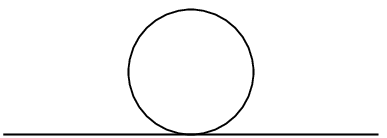}}}
\Rightarrow  m_H^2 \propto \Lambda^2~,
\eeq
leading to the hierarchy and naturalness problems.\cite{thooft} 
Finally, the $\beta$ function for the self-coupling $\lambda$
\beq
{\lower7pt\hbox{\epsfysize=0.25 truein \epsfbox{beta.eps}}}
\Rightarrow \beta = {3\lambda^2 \over 2 \pi^2} \, > \, 0
{}~,
\eeq
leading to a ``Landau pole'' and triviality.\cite{trivial}

\begin{figure}[tbp]
\centering
\epsfysize=2in
\hspace*{0in}
\epsffile{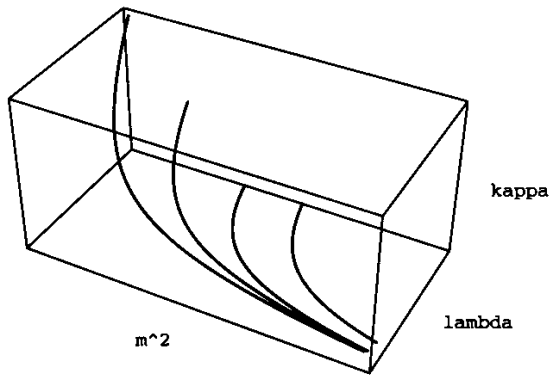}
\caption{Renormalization group flow of Higgs mass $m^2$, Higgs
self-coupling $\lambda$, and the coefficient of a representative
irrelevant operator $\kappa$. The flows go from upper-left to
lower-right as one scales to the infrared.}
\label{Fig1}
\end{figure}

The hierarchy/naturalness and triviality problems
can be nicely summarized in terms of the 
Wilson renormalization group.
Define the theory with a fixed UV-cutoff:
\beqa
{\cal L}_\Lambda =  & D^\mu \phi^\dagger D_\mu \phi + 
m^2(\Lambda)\phi^\dagger \phi 
+ {\lambda(\Lambda)\over 4}(\phi^\dagger\phi)^2 \\
& + {\kappa(\Lambda)\over 36\Lambda^2}(\phi^\dagger\phi)^3+\ldots  
\nonumber 
\eeqa
Here $\kappa$ is the coefficient of a representative 
irrelevant operator, 
of dimension greater than four.
Next, integrate out states with $\Lambda^\prime < k < \Lambda$,
and construct a new Lagrangian with the same {\it
low-energy} Green's functions:
\beqa
{\cal L}_\Lambda & \Rightarrow & {\cal L}_{\Lambda^\prime} \nonumber\\
m^2(\Lambda)& \rightarrow & m^2(\Lambda^\prime) \nonumber \\
\lambda(\Lambda) & \rightarrow & \lambda(\Lambda^\prime) \nonumber \\
\kappa(\Lambda) & \rightarrow & \kappa(\Lambda^\prime)  
\eeqa
The low-energy behavior of the theory is then nicely summarized in terms
of the evolution of couplings in the infrared.\footnote{For convenience,
  we ignore the corrections due to the weak gauge interactions.  In
  perturbation theory, at least, the presence of these interactions does
  not qualitatively change the features of the Higgs sector.} A
three-dimensional representation of this flow in the
infinite-dimensional space of couplings shown in Figure \ref{Fig1}.

From Figure \ref{Fig1}, we see that as we scale to the infrared the
coefficients of irrelevant operators, such as $\kappa$, tend zero; {\it
  i.e.} the flows are attracted to the finite dimensional subspace
spanned (in perturbation theory) by operators of dimension four or less; this
is the modern understanding of {\it renormalizability}. On the other
hand, the coefficient of the only {\it relevant} operator (of dimension
2), $m^2$, tends to infinity. This leads to the
naturalness \& hierarchy problem.\cite{thooft} Since we want $m^2 \propto
v^2$ at low energies we must adjust the value of $m^2(\Lambda)$
to a precision of 
\beq
{\Delta m^2(\Lambda) \over m^2(\Lambda)} \propto {v^2 \over \Lambda^2}~.
\eeq
Finally, the coefficient of the only marginal operator
$\lambda$ tends, because of the positive $\beta$ function, 0.
If we try to take the continuum limit, $\Lambda \to +\infty$,
the theory becomes free or trivial.\cite{trivial} This last statement
implies that, in and of itself, the standard one-doublet higgs
model is incomplete.

The analysis we have presented is based on perturbation theory and is
valid in the domain of attraction of the ``Gaussian fixed point''
($\lambda=0$).  In principle, however, the Wilson approach can be used
{\it non-perturbatively} and take into account the presence of
nontrivial fixed points or large anomalous dimensions.  In a
conventional Higgs theory, neither of these effects is thought to 
occur~\cite{lattice}
--- these issues will, however, be relevant in theories of dynamical
electroweak symmetry breaking.

\subsection{Solving the Naturalness/Hierarchy Problems}

There are only two ways of dealing with any hierarchy,
political or otherwise: we can either stabilize or
eliminate it.

The conservative approach of stabilizing the hierarchy can be
implemented by introducing a symmetry which protects the scalar masses.
One approach is supersymmetry.\cite{giudice} In this case each scalar
is associated with a fermionic superpartner and the chiral symmetry of the
superpartners of the scalar higgs protects the mass from receiving
corrections of ${\cal O}(\Lambda^2)$. In practice this
occurs because of a cancelation between loop-diagrams
involving scalars and fermions, for example
\beq
\hskip-15pt{\lower20pt\hbox{
\epsfysize=0.5 truein \epsfbox{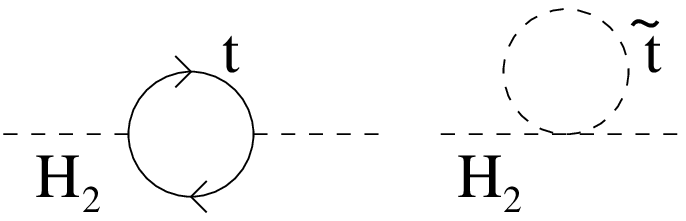}}
\rightarrow \delta m_H^2 \propto \log \Lambda^2}~.
\eeq
An alternative approach to stabilizing the hierarchy is to
use the ``composite higgs'' approach of Georgi and Kaplan.\cite{composite}
In these models, the higgs is a Goldstone boson whose
mass is protected by a (spontaneously broken) chiral symmetry. In
these models electroweak symmetry breaking is due to 
``vacuum (mis)-alignment.''

Models of dynamical electroweak symmetry are based on the radical
approach of eliminating the hierarchy. Here electroweak symmetry
breaking is due to {\it chiral symmetry breaking} in a gauge theory with
massless fermions. We will concentrate on this approach in
what follows.

\subsection{Technicolor: A Dynamical Electroweak Symmetry Breaking}

The simplest theory of dynamical electroweak symmetry
breaking is technicolor.\cite{tc} Consider
an $SU(N_{TC})$ gauge theory with fermions in the fundamental
representation of the gauge group
\beq
\Psi_L=\left(
\begin{array}{c}
U\\D
\end{array}
\right) _L\,\,\,\,\,\,\,\,
U_R,D_R
\eeq
The fermion kinetic energy terms
for this theory are
\beqa
{\cal L} &=& \bar{U}_L i\slashchar{D} U_L+
\bar{U}_R i\slashchar{D} U_R+\\
 & &\bar{D}_L i\slashchar{D} D_L+
\bar{D}_R i\slashchar{D} D_R~,
\nonumber
\eeqa
and, like QCD in $m_u$, $m_d \to 0$ limit, have
a chiral $SU(2)_L \times SU(2)_R$ symmetry.

As in QCD, exchange of technigluons in the spin zero,
isospin zero channel is attractive
\beq
\hskip-10pt{\lower15pt\hbox{
\epsfysize=0.5 truein \epsfbox{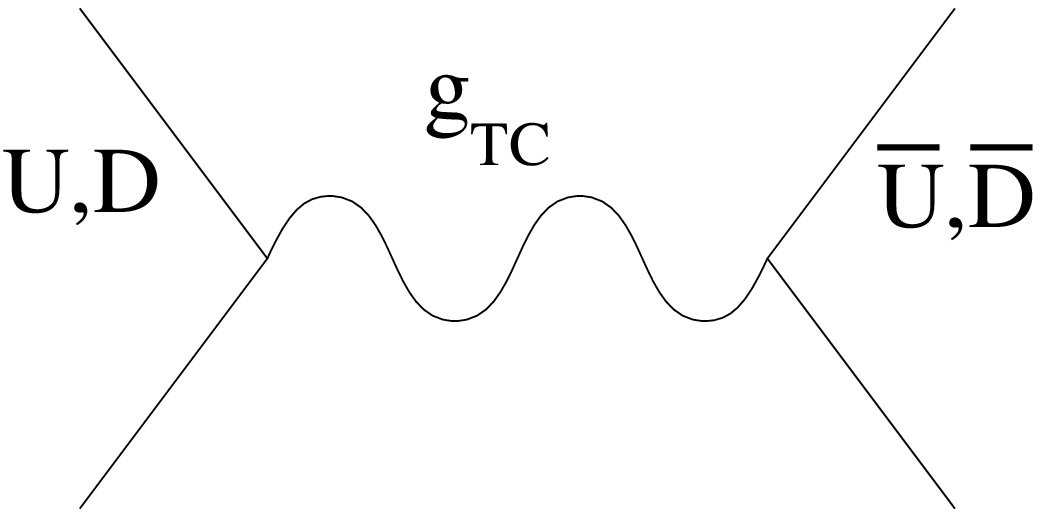}}
\rightarrow  \langle \bar U_LU_R\rangle
=\langle \bar D_LD_R\rangle \neq 0\, ,}
\eeq
causing the formation of a condensate which dynamically breaks
$SU(2)_L \times SU(2)_R \to SU(2)_V$.
These broken chiral symmetries imply the existence
of three massless Goldstone bosons, the analogs of the
pions in QCD.

Now consider gauging $SU(2)_W \times U(1)_Y$ with the left-handed
fermions transforming as weak doublets and the right-handed ones as weak
singlets (in this one-doublet model we will take the left-handed
technifermions to have hypercharge zero and the right-handed up- and
down-technifermions to have hypercharge $\pm 1/2$).  The spontaneous 
breaking of the chiral symmetry breaks the weak-interactions
down to electromagnetism. The would-be Goldstone bosons become
\beq
\pi^\pm,\, \pi^0 \, \rightarrow\, W^\pm_L,\, Z_L~,
\eeq
the longitudinal components of the $W$ and $Z$ which
acquire a mass
\beq
M_W = {g F_{TC} \over 2}~.
\eeq
Here $F_{TC}$ is the analog of $f_\pi$ in QCD. In order
to obtain the experimentally observed masses, we must have
that $F_{TC} \approx 250 {\rm GeV}$ and hence this model
is essentially QCD scaled up by a factor of
\beq
{F_{TC}\over f_\pi} \approx 2500\, .
\eeq

While I have described only the simplest model above, it is
straightforward to generalize to other cases.  {\it Any} strongly
interacting gauge theory with a chiral symmetry breaking pattern $G \to
H$, in which $G \supset SU(2)_W \times U(1)_Y $ and breaks to a
subgroup $H \supset U(1)_{em}$ (with $SU(2)_W \times U(1)_Y \not\subset
H$) will break the weak interactions down to electromagnetism.
In order to correspond to experimental results, however, we must
also require that $H$ contain ``custodial'' $SU(2)_C$ which
insures that the $F$-constant associated with the $W^\pm$ and
$Z$ are equal and therefore that the relation
\beq
\rho = {M_W \over M_Z \sin\theta_W} =1
\eeq
is satisfied at tree-level.  If the chiral symmetry
is larger than $SU(2)_L\times SU(2)_R$, theories of this sort will
contain additional (pseudo-)Goldstone bosons which are not ``eaten'' by
the $W$ and $Z$.  For simplicity, in the remainder of this lecture, we
will discuss the phenomenology of the one-doublet model.\cite{dpf}

\subsection{The Phenomenology of Dynamical Electroweak Symmetry Breaking}

Of the particles that we have observed to date, the only ones directly
related to the electroweak symmetry breaking sector~\footnote{Except,
  possibly, for the third generation. See the discussion of topcolor in
  lecture 3.} are the longitudinal gauge-bosons. Therefore, we expect
that the most direct signatures for electroweak symmetry breaking to
come from the scattering of longitudinally gauge bosons. At {\it
  high-energies}, we may use the equivalence theorem~\cite{equiv}
\beq
{\cal A}(W_L W_L) = {\cal A}(\pi \pi) + {\cal O}({M_W\over E})\, .
\eeq
to reduce the problem of longitudinal gauge boson ($W_L$) scattering to
the corresponding (and generally simpler) problem of the scattering of
the Goldstone bosons ($\pi$) that would be present in the absence of
the weak gauge interactions.

In order to correctly describe the weak interactions,
the symmetry breaking sector must have an (at least approximate)
custodial symmetry,\cite{custodial} and the most general
effective theory describing the behavior of the Goldstone
bosons is an effective chiral lagrangian~\cite{chiral}
with an $SU(2)_L \times SU(2)_R \to SU(2)_V$ symmetry
breaking pattern. This effective lagrangian
is most easily written in terms of a field
\beq
\Sigma = \exp(i\pi^a\sigma^a/F_{TC})~,
\eeq
where the $\pi^a$ are the Goldstone boson
fields, the $\sigma^a$ are the Pauli matrices, and
where the field $\Sigma$ which transforms as 
\beq
\Sigma  \to L \Sigma R^\dagger
\eeq
under $SU(2)_L \times SU(2)_R$.

The interactions can then be ordered in a power-series in
momenta.  Allowing for custodial $SU(2)$ violation, the
lowest-order terms in the effective theory are
\beq
{   {F_{TC}^2 \over4} \tr\left[D^\mu
   \Sigma^{\dagger}D_\mu \Sigma\right] +  {F_{TC}^2 \over 2} ({1\over\rho} - 1) 
\left[ \tr T_3 \Sigma^\dagger D^\mu \Sigma \right]^2}
\label{psquare}
\eeq
where
\beq
D_\mu \Sigma=\partial_\mu \Sigma+\im g\W_\mu 
\Sigma-\im \Sigma g'\B_\mu~,
\eeq
and the gauge-boson kinetic terms
\beq
 -{1\over2}\tr\left[\W^{\mu\nu}\W_{\mu\nu}\right]- 
 {1\over2}\tr\left[\B^{\mu\nu} \B_{\mu\nu}\right]\, .
\eeq
In unitary gauge, $\Sigma=1$ and the lowest-order
terms in eqn. \ref{psquare} give rise to the
$W$ and $Z$ masses
\beq
{g^2 F_{TC}^2\over 4} W^{-\mu} W^+_{\mu} + {g^2 F_{TC}^2\over{8 \rho \cos^2\theta}}
Z^\mu Z_\mu
~.
\eeq

So far, the description we have constructed is valid in {\it any} theory
of electroweak symmetry breaking.  The interactions in eqn.
\ref{psquare} result in {\it universal} low-energy~\cite{let}
theorems
\beqa 
{\cal M}[W^+_L W^-_L \to W^+_L W^-_L] =& {i u \over {v^2 \rho}}  \nonumber\\
{\cal M}[W^+_L W^-_L \to Z_L Z_L] =& {i s \over {v^2}} \left( 4 -
{3\over\rho}\right) \\
{\cal M}[Z_L Z_L \to Z_L Z_L] =& 0~. \nonumber
\eeqa
These amplitudes increase with energy and, at some
point, this growth must stop.\cite{lqt} 
What dynamics cuts off growth in these amplitudes?
In general, there are three possibilities:
\begin{itemize}
\item New particles
\item The born approximation fails $\rightarrow$ strong interactions
\item both.
\end{itemize}

\begin{figure}[tbp]
\centering
\epsfysize=2in
\hspace*{0in}
\epsffile{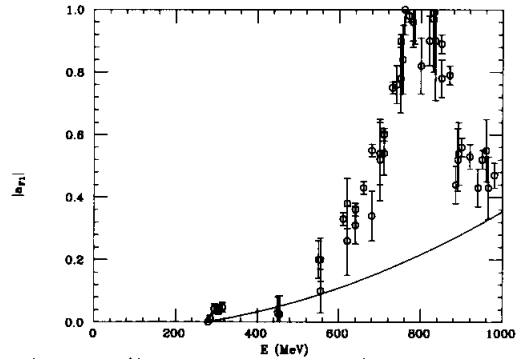}
\caption{QCD data~\protect\cite{donoghue}and low-energy theorem
(solid line) prediction for the magnitude of the spin-1/isospin-1 
pion scattering amplitude $|a_{F1}|$.}
\label{Fig2}
\end{figure}

In the case of QCD-like technicolor, we take our inspiration
from the familiar strong interactions. The
data for $\pi\pi$ scattering in QCD in the
$I=J=1$ channel is shown in Figure \ref{Fig2}.
After correcting for the finite pion mass, we see that
the scattering amplitude follows the low-energy prediction
near threshold, but at higher energies the amplitude is
dominated by the $\rho$-meson whose appearance (1)
enhances the scattering cross-section and (2) cuts-off
the growth of the scattering amplitude at higher energies.
In a QCD-like technicolor theory, then, we expect
the appearance of a vector meson whose mass we
estimate by scaling by $F_{TC}/f_\pi \approx 2500$. That is,
\beq
M_{\rho_{TC}} \approx 2 \tev \, \sqrt{3\over N_{TC}}~,
\eeq
where we have included large-$N_{TC}$ scaling to estimate the
effect of $N_{TC} \neq 3$.\cite{largeN}

\begin{figure}[tbp]
\centering
\hskip-5pt\hbox{
{\epsfxsize=4.5cm \epsfbox{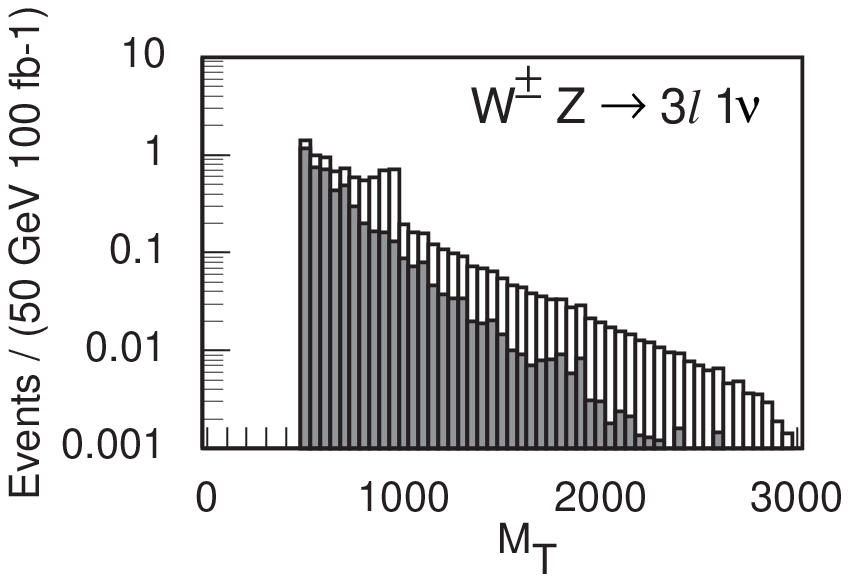}}
\hskip10pt
{\epsfxsize=4.5cm \epsfbox{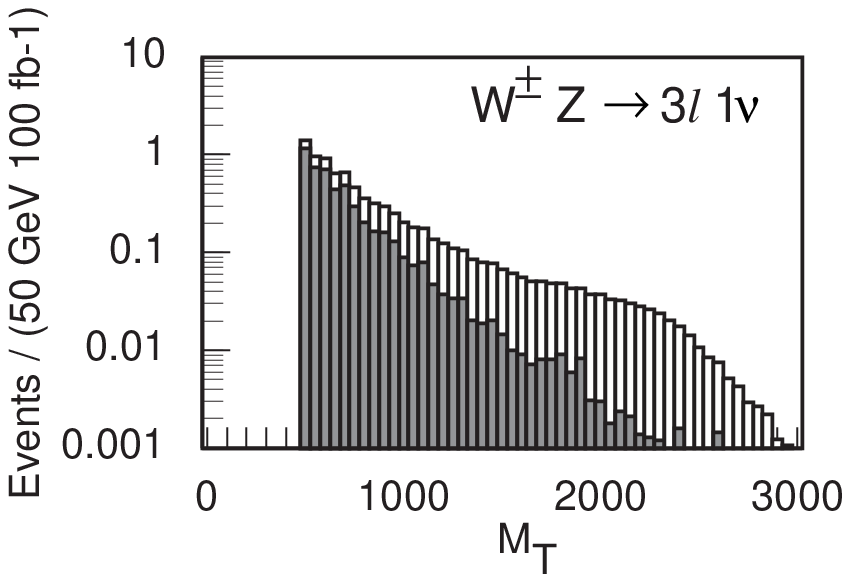}}
}
{\lower 25pt\hbox{\epsfxsize=5cm
\epsfbox{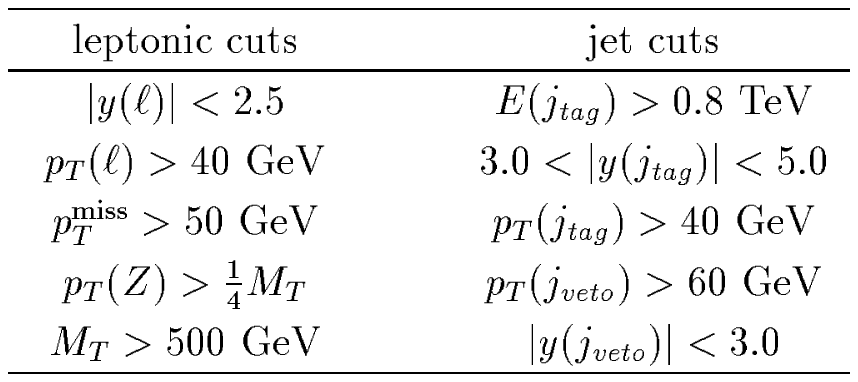}}}
\caption{Gauge boson scattering signal plus background (grey) and background (black) for
$W^\pm Z$ production~\protect\cite{bagger} at LHC for technirho masses of 1.0 TeV and
2.5 TeV. Signal selection requirements shown in table above.}
\label{Fig3}
\end{figure}

The most direct experimental signature of dynamical
electroweak symmetry breaking is to look for these
``technivector mesons.'' At the LHC, gauge boson
scattering occurs through the following process,
\beq
{\lower20pt\hbox{{\epsfxsize=2cm \epsfbox{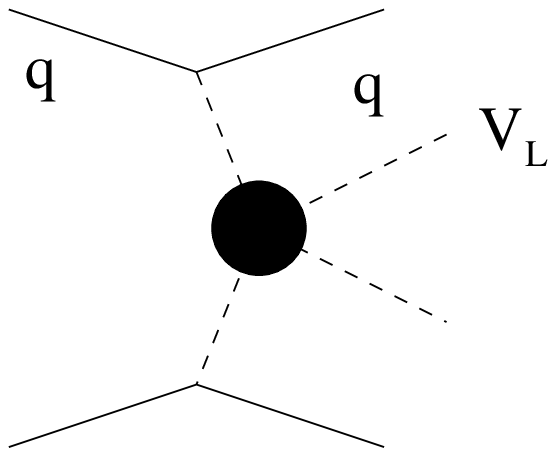}}}}~.
\eeq
Note that, in addition to high-$p_T$ gauge bosons, one
expects forward ``tag'' jets (with a typical transverse
momentum of order $M_W$) from the quarks which radiate
the initial gauge bosons.
The signal expected is shown~\cite{bagger} in Figure \ref{Fig3}
for $M_{\rho_{TC}}= 1.0\,\tev, 2.5\,\tev$. Note the
scale: events per 50 GeV bin of transverse mass ($M_T$)
per 100 fb$^{-1}$!

\begin{figure}[tbp]
\centering
\hskip5cm\hbox{\epsfxsize=8cm\epsfbox{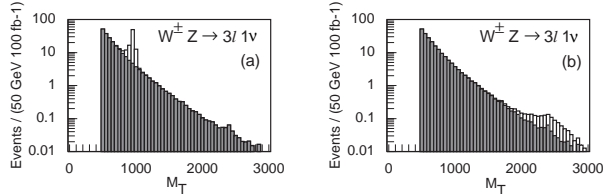}}
\caption{Vector meson mixing signal plus background (grey) and background (black) for
$W^\pm Z$ production~\protect\cite{golden} at LHC for technirho masses of (a) 1.0 TeV and
(b) 2.5 TeV.}
\label{Fig4}
\end{figure}

A complementary signal is provided through the technicolor
analog of ``vector-meson dominance.'' In particular, the
$W$ and $Z$ can mix with the technirho mesons in a manner
exactly analogous to $\gamma$-$\rho$ mixing in QCD:
\beq
{\lower5pt\hbox{{\epsfxsize=2cm \epsfbox{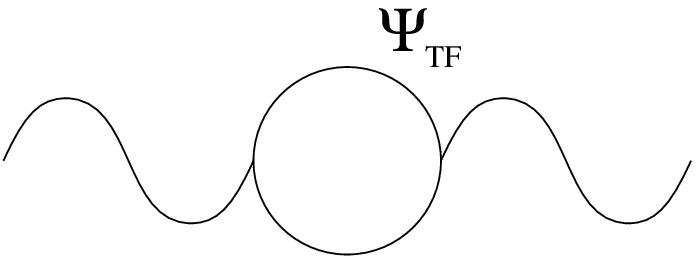}}}}
\hskip5pt 
\Rightarrow
\hskip5pt
{\lower15pt\hbox{{\epsfxsize=2cm \epsfbox{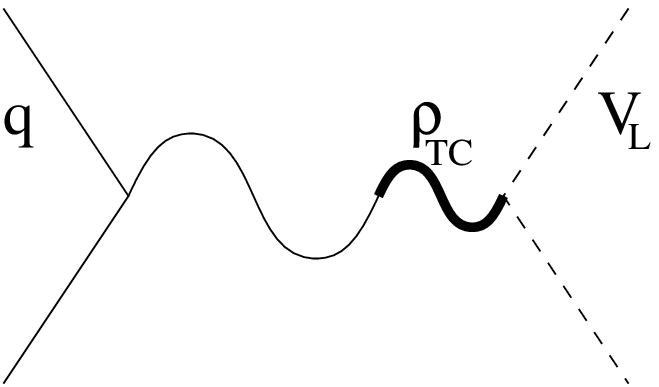}}}}~.
\eeq
Note that this process does {\it not} have a very
forward jet and is distinguishable from the gauge boson
scattering signal discussed above.
The vector-meson mixing signal~\cite{golden} at the LHC is shown in
Figure \ref{Fig4} for $M_{\rho_{TC}}= 1.0\,\tev, 2.5\,\tev$.

\begin{figure}[tbp]
\centering
\hskip0.3cm\hbox{\epsfxsize=5cm\epsfbox{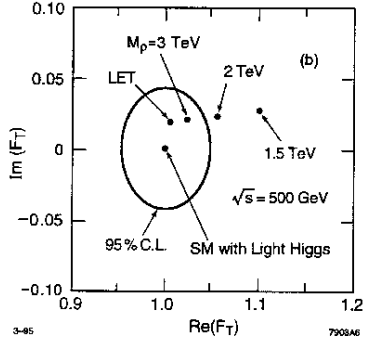}}
\caption{$ZWW$ form-factor measurement~\protect\cite{barklow} 
at a 500 GeV NLC with 80fb$^{-1}$. Predictions are shown for the standard model,
and for technicolor for various technirho masses.}
\label{Fig5}
\end{figure}

A dynamical electroweak symmetry breaking sector will
also have effect two gauge-boson production at a
high-energy $e^+ e^-$ collider such as the NLC.
For example, if gauge-boson re-scattering~\footnote{If
the technicolor theory satisfies a ``KSRF'' relation,\cite{ksrf}
this ``re-scattering'' effect is exactly equivalent to the
vector-meson mixing effect discussed above.\cite{kroll}}
\beq
{\lower20pt\hbox{{\epsfxsize=3cm \epsfbox{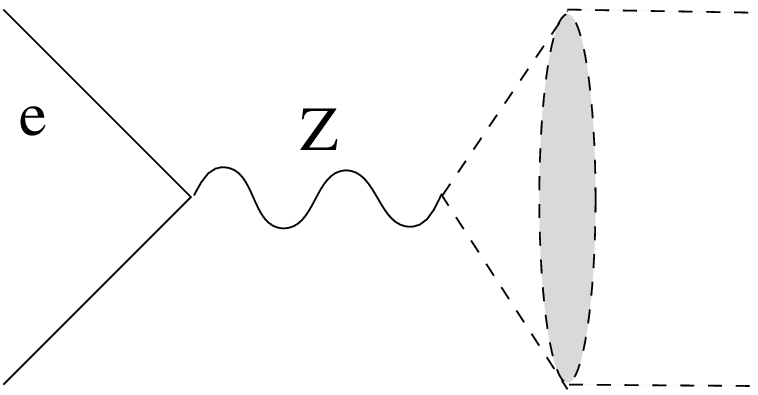}}}}
\eeq
is dominated by a technirho meson, it 
can be parameterized in terms of a $ZWW$ form-factor
\beq
 F_T =
         \exp\bigl[{1\over \pi} \int_0^\infty
          ds'\delta(s',M_\rho,\Gamma_\rho)
          \{ {1\over s'-s-i\epsilon}-{1\over s'}\}
         \bigr]~,
\eeq
where
\beq
\delta(s) = {1\over 96\pi} {s\over v^2}
+ {3\pi\over 8} \left[ \tanh (
{
s-M_\rho^2
\over
M_\rho\Gamma_\rho
}
)+1\right]~.
\eeq
This two gauge-boson production mechanism interferes with
continuum production, and by an accurate measurement of 
the decay products is is possible~\cite{barklow} to reconstruct
the real and imaginary parts of the form-factor $F_T$.
The expected accuracy of a 500 GeV NLC with 80 fb$^{-1}$ is
shown in Figure \ref{Fig5}.

\subsection{Low-Energy Phenomenology}

Even though the most direct signals of a dynamical electroweak symmetry
breaking sector require (partonic) energies of order 1 TeV, there are
also effects which may show up at lower energies as well. While the
${\cal O}(p^2)$ terms in the effective lagrangian are universal, terms of
higher order are model-dependent.  At energies below $M_{\rho_{TC}}$,
there are corrections to 3-pt functions:
\beq
{\lower15pt\hbox{\epsfysize=0.75 truein \epsfbox{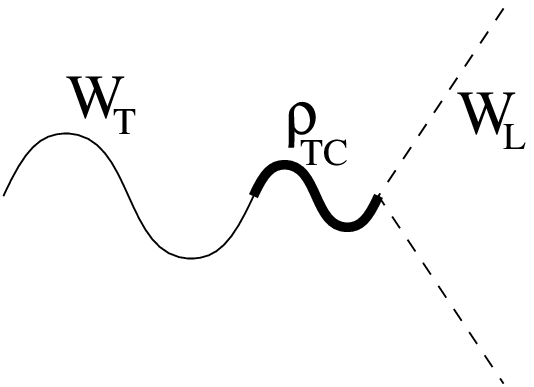}}}~,
\eeq
which, following Gasser and Leutwyler,\cite{chiral} give
rise to the {\cal O}($p^4$) terms
\beq
 -\ i g {{\it l}_{9L} \over 16 \pi^2}\, \tr {\W^{\mu \nu} D_\mu
\Sigma D_\nu \Sigma^\dagger}~,
\eeq
and
\beq
-\ i g' {{\it l}_{9R} \over 16 \pi^2}\, \tr {\B^{\mu \nu}
D_\mu \Sigma^\dagger D_\nu\Sigma}~,
\eeq
as well as corrections to the 2-pt functions:
\beq
{\lower5pt\hbox{\epsfysize=0.3 truein \epsfbox{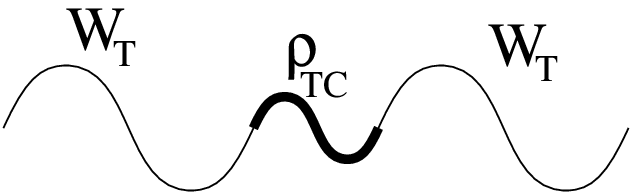}}}
\eeq
which gives rise to
\beq
\  +\ g g' {{\it l}_{10}\over 16 \pi^2}\, \tr {\Sigma \B^{\mu \nu}
\Sigma^\dagger \W_{\mu \nu}}
~.
\eeq
In these expressions, the {\it l}'s are normalized to be {\cal O}(1).

The corrections to the 3-point functions are typical,
following Hagiwara, {\it et. al.},\cite{hagiwara}
parameterized:
\beqa
{i\over e \cot\theta} & {\cal L}_{WWZ}  =  g_1
(W^\dagger_{\mu\nu} W^\mu Z^\nu - W^\dagger_\mu Z_\nu W^{\mu\nu}) \nonumber \\
& + \kappa_Z W^\dagger_\mu W_\nu Z^\mu\nu + {\lambda_Z\over
M_W^2}W^\dagger_{\lambda\mu}W^\mu_\nu Z^{\nu\lambda}~,
\eeqa
and
\beqa
{i\over e}  & {\cal L_{WW\gamma}} =  (W^\dagger_{\mu\nu} W^\mu A^\nu -
W_\mu^\dagger A_\nu W^{\mu\nu})\nonumber \\
&  + \kappa_\gamma W^\dagger_\mu W_\nu
F^\mu\nu + {\lambda_\gamma\over M_W^2} W^\dagger_{\lambda\mu}W^\mu_\nu
F^{\nu\lambda}~.
\eeqa
Re-expressing these coefficients in terms of 
the parameters in ${\cal L}_{p^4}$ given above, we find
\beq
\left.
\begin{array}{c}
g_1 - 1\\
\kappa_Z - 1\\
\kappa_\gamma - 1
\end{array}
\right\}
\ \approx\ {\alpha_* l_i \over 4\pi \sin^2\theta} = {\cal O}(10^{-2}-10^{-3})~,
\eeq
and $\lambda_{Z,\gamma}$ from ${\cal L}_{p^6}$ implying that
\beq
\lambda_{Z,\gamma} = {\cal O}(10^{-4}-10^{-5})\, .
\eeq

\begin{figure}[tbp]
\centering
\hskip-5pt\hbox{
{\epsfxsize=4.25cm \epsfbox{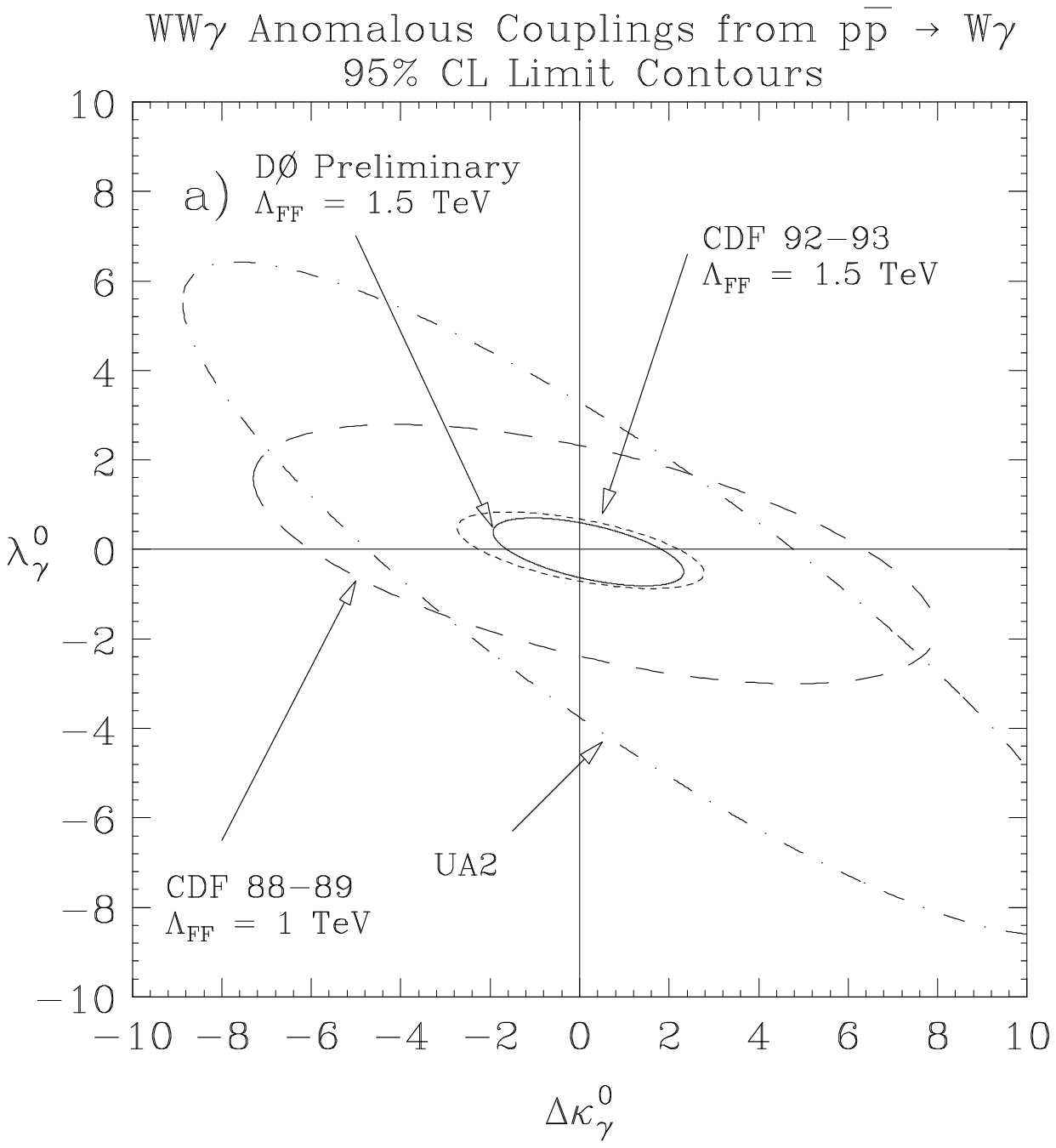}}
\hskip10pt
{\epsfxsize=4.25cm \epsfbox{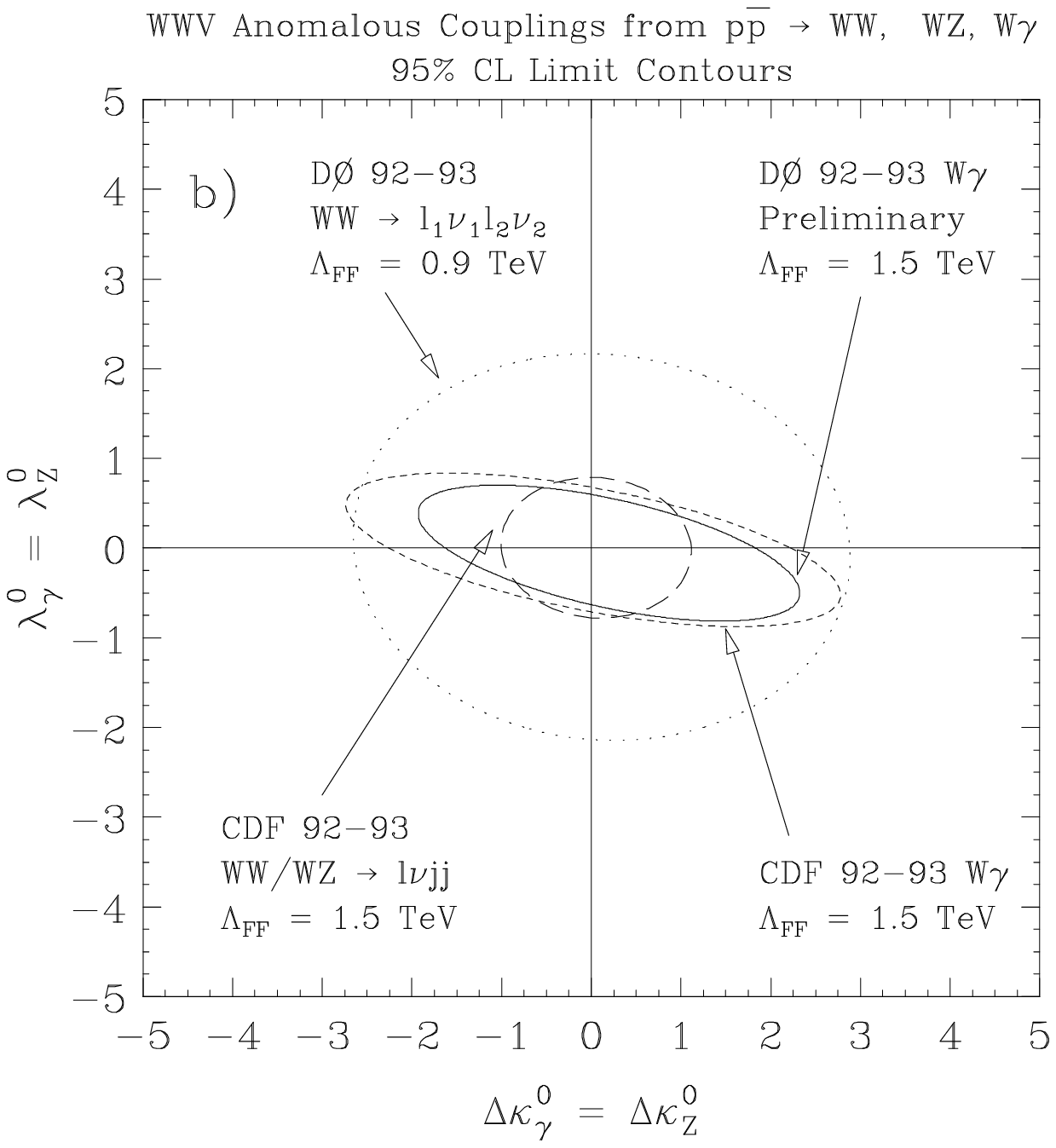}}
}
\caption{Current limits~\protect\cite{aihara} on anomalous 
gauge-boson vertices from Tevatron data.}
\label{Fig6}
\end{figure}

\begin{figure}[tbp]
\centering
\epsfxsize 10cm \centerline{\epsffile{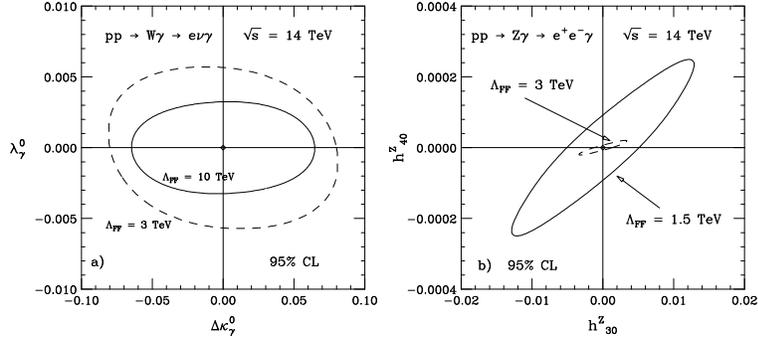}}
\caption{Experimental~\protect\cite{aihara} reach of LHC to probe
anomalous gauge-boson vertices given an integrated luminosity
of 100 fb$^{-1}$.}
\label{Fig7}
\end{figure}

\begin{figure}[tbp]
\centering
\epsfxsize 4.5cm \centerline{\epsffile{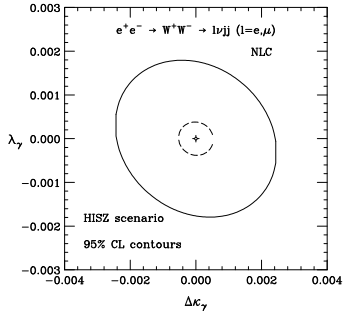}}
\caption{Experimental~\protect\cite{aihara} reach of a
500 GeV (solid) or 1500 GeV (dashed) NLC to probe
anomalous gauge-boson vertices, assuming 80 fb$^{-1}$ or
190 fb$^{-1}$ respectively.}
\label{Fig8}
\end{figure}

The best current limits,\cite{aihara} coming from Tevatron experiments
are shown in Figure \ref{Fig6}. Unfortunately, they do not reach the
level of sensitivity required. The situation~\cite{aihara} is somewhat
improved at the LHC, as shown in Figure \ref{Fig7}, or at a 500 or 1500
GeV NLC, as shown in Figure \ref{Fig8}.

\begin{figure}[tbp]
\centering
\epsfxsize 8cm \centerline{\epsffile{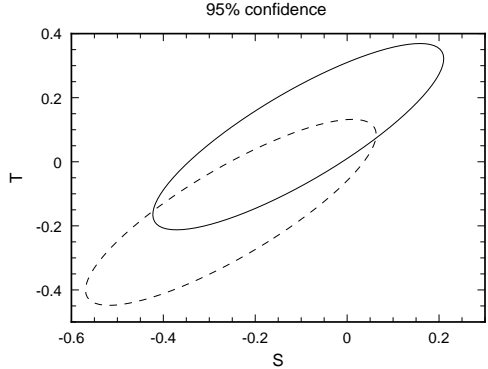}}
\caption{95\% confidence level bounds~\protect\cite{terning} on
the oblique parameters $S$ and $T$ for 
$\alpha_S = 0.115$ (solid) and 0.124 (dashed).}
\label{Fig9}
\end{figure}

The corrections~\cite{oblique} to the 2-pt functions give rise to contributions
to the ``oblique parameters'' $S$ 
\beqa
&S&\equiv 16\pi \left[ \Pi'_{33}(0) - \Pi'_{3Q}(0)\right] \nonumber\\
&=& -\pi {\it l}_{10}\approx
4\pi\left({F^2_{\rho_{TC}}\over M^2_{\rho_{TC}}}
-{F^2_{A_{TC}}\over M^2_{A_{TC}}}\right) N_D~,
\eeqa
and $T$
\beq
\alpha T \equiv {g^2 \over {\cos^2\theta M_Z^2}} \left[\Pi_{11}(0) -
\Pi_{33}(0)\right] = \rho-1~.
\eeq
Current experimental constraints~\cite{terning} imply the
bounds shown in Figure \ref{Fig9}, at 95\% confidence
level for different values of $\alpha_S$. Scaling from
QCD, we expect a contribution to S of order 
\beq
S \approx 0.28 N_D (N_{TC}/3)~,
\eeq
for an $SU(N_{TC})$ technicolor theory with $N_D$ technidoublets.
From these we see that, with the possible exception of 
$N_D=1$ and $N_{TC}=2$ or 3, {\it QCD-like} technicolor is in
conflict with precision weak measurements.

Dynamical Electroweak Symmetry Breaking provides a natural and
attractive mechanism for producing the $W$ and $Z$ masses. {\it
  Generically} models of this type predict strong $WW$-Scattering,
signals of which may be observable at the LHC. While the simplest
QCD-like models serve as a useful starting point, they are excluded
(except, perhaps, for an $SU(2)_{TC}$ model with one doublet) since they
would give rise to unacceptably large contributions to the $S$
parameter. In the next lecture we will discuss the additional
interactions and features required in (more) realistic models to give
rise to the masses to the ordinary fermions.

\section{Lecture 2: Flavor Symmetry Breaking and ETC}

\subsection{Fermion Masses \& ETC Interactions}

In order to give rise to masses for the ordinary quarks
and leptons, we must introduce interactions which
connect the chiral-symmetries of technifermions
to those of the ordinary fermions. The most popular
choice~\cite{DimSuss,EL} is to introduce new broken gauge interactions, called
{\it extended technicolor interactions} (ETC), which couple 
technifermions to ordinary fermions. At energies low compared
to the ETC gauge-boson mass, $M_{ETC}$, these effects
can be treated as local four-fermion interactions
\beq
{\lower15pt\hbox{\epsfysize=0.5 truein \epsfbox{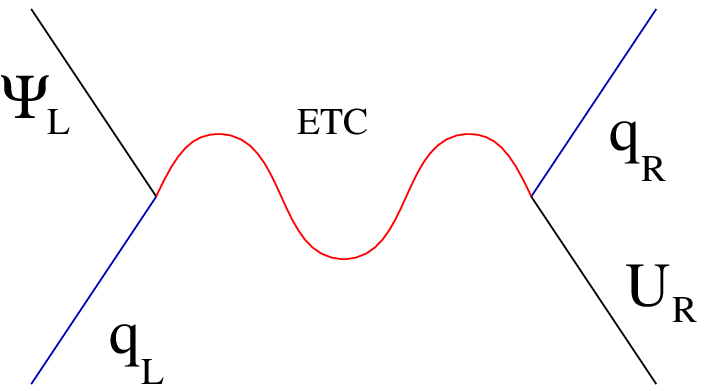}}}
\Rightarrow {{g_{ETC}^2\over M^2_{ETC}}}(\overline{\Psi}_L U_R)
({\overline{q}_R q_L})~.
\label{etcint}
\eeq
After technicolor chiral-symmetry breaking, such an
interaction gives rise to a mass for an ordinary fermion
\beq
m_q \approx {{g_{ETC}^2\over M^2_{ETC}}} \langle\overline{U} U\rangle_{ETC}~,
\label{fmass}
\eeq
where $\langle \overline{U} U\rangle_{ETC}$ is the value of the
technifermion condensate evaluated at the ETC scale (of order
$M_{ETC}$).  The condensate renormalized at the ETC scale in eqn.
\ref{fmass} can be related to the condensate renormalized at the
technicolor scale as follows
\beq
\langle\overline{U} U\rangle_{ETC} = \langle\overline{U} U\rangle_{TC}
\exp\left(\int_{\Lambda_{TC}}^{M_{ETC}} {d\mu \over \mu}
\gamma_m(\mu)\right)~,
\eeq
where $\gamma_m(\mu)$ is the anomalous dimension of the
fermion mass operator and $\Lambda_{TC}$ is the analog of $\Lambda_{QCD}$
for the technicolor interactions.

For QCD-like technicolor (or any theory which
is ``precociously'' asymptotically free), $\gamma_m$ is
small over in the range between $\Lambda_{TC}$ and $M_{ETC}$ and
using dimensional analysis~\cite{chiral} we find
\beq
\langle\overline{U} U\rangle_{ETC} \approx \langle\overline{U} U\rangle_{TC}
\approx 4\pi F^3_{TC}~.
\eeq
In this case eqn. \ref{fmass} implies that
\beq
{ {M_{ETC}\over g_{ETC}}} \approx 40 \tev 
\left({F_{TC}\over 250\gev}\right)^{3\over 2}
\left({100 \mev \over m_q}\right)^{1\over 2}~.
\eeq

In order to orient our thinking, it is instructive to consider a simple
``toy'' extended technicolor model. The model is based on an
$SU(N_{ETC})$ gauge group, with technicolor as an extension of flavor.
In this case $N_{ETC} = N_{TC} + N_F$, and we add the (anomaly-free)
set of fermions
\medskip
\begin{center}
$
\begin{array}{ll}
Q_L=(N_{ETC},3,2)_{1/6} & L_L=(N_{ETC},1,2)_{-1/2} \\
U_R=(N_{ETC},3,1)_{2/3} & E_R=(N_{ETC},1,1)_{-1} \\
D_R=(N_{ETC},3,1)_{-1/3} & N_R=(N_{ETC},1,1)_{0}~,
\end{array}
$
\end{center}
\medskip\noindent
where we display their quantum numbers under $SU(N_{ETC})\times
SU(3)_C \times SU(2)_W \times U(1)_Y$. We break the
ETC group down to technicolor in three stages
\medskip
\begin{center}
{$SU(N_{TC}+3)$}
\end{center}
\begin{center}
$\Lambda_1\ \ \ \ \ \downarrow \ \ \ \ \  
m_1\approx{4\pi F^3\over \Lambda^2_1}$
\end{center}
\begin{center}
{$SU(N_{TC}+2)$}
\end{center}
\begin{center}
$\Lambda_2\ \ \ \ \ \downarrow \ \ \ \ \  
m_2\approx{4\pi F^3\over \Lambda^2_2}$
\end{center}
\begin{center}
{$SU(N_{TC}+1)$}
\end{center}
\begin{center}
$\Lambda_3\ \ \ \ \ \downarrow \ \ \ \ \  
m_3\approx{4\pi F^3\over \Lambda^2_3}$
\end{center}
\begin{center}
{$SU(N_{TC})$}
\end{center}
\medskip\noindent
resulting in three isospin-symmetric families of degenerate
quarks and leptons, with $m_1 < m_2 < m_3$. Note that the
{\it heaviest} family is related to the {\it lightest} ETC 
scale!

Before continuing our general discussion, it is worth noting
a couple of points. First,
in this example the ETC gauge-boson do not carry color
or weak charge
\beq
[G_{ETC},SU(3)_C]=[G_{ETC},SU(2)_W]=0~.
\label{commute}
\eeq
Furthermore, in this model there is one technifermion for each type of
ordinary fermion: that is, this is a ``one-family'' technicolor
model.\cite{farhi}  Since there are eight left- and right- handed
technifermions, the chiral symmetry of the technicolor theory is (in the
limit of zero qcd and weak couplings) $SU(8)_L \times SU(8)_R \to
SU(8)_V$. Such a theory would lead to $8^2-1=63$ (pseudo-)Goldstone
bosons. Three of these Goldstone bosons are unphysical --- the
corresponding degrees of freedom become the longitudinal components of
the $W^\pm$ and $Z$ by the Higgs mechanism.  The remaining 60 must
obtain a mass, and the condition in eqn. \ref{commute} will be modified
in a realistic model.  We will return to the issue of pseudo-Goldstone
bosons below.

The most important feature of this or any ETC-model is that a successful
extended technicolor model will provide a {\it dynamical theory of
  flavor}! As in the toy model described above and as explicitly
shown in eqn. \ref{etcint} above, the masses of the ordinary fermions
are related to the masses and couplings of the ETC gauge-bosons. A successful
and complete ETC theory would predict these quantities, and hence the
ordinary fermion masses. 

Needless to say, constructing such a theory is very difficult. No
complete \& successful theory has been proposed.  Examining our toy
model, we immediately see a number of shortcomings of this model that
will have to be addressed in a more realistic theory:
\begin{itemize}
\item What breaks ETC?
\item Do we require a { separate} scale for each family?
\item How do we obtain quark mixing angles?
\item $T_3 = \pm {1\over 2}$ fermions have { equal} masses, hence the $u_R$ 
\& $d_R$ must be in different representations~\cite{EL} of ETC.
\item What about right-handed technineutrinos and $m_\nu$?
\end{itemize}

\subsection{Flavor-Changing Neutral-Currents}

Perhaps the single biggest obstacle to constructing a realistic ETC
model (or any dyanmical theory of flavor) is the potential for
flavor-changing neutral currents.\cite{EL}  Quark mixing implies
transitions between different generations: $q \to \Psi \to q^\prime$,
where $q$ and $q'$ are quarks of the same charge from different
generations and $\Psi$ is a technifermion. Consider the commutator of
two ETC gauge currents:
\beq
[\overline{q}\gamma \Psi, \overline{\Psi}\gamma q^\prime] \supset 
\overline{q}\gamma q^\prime\, .
\eeq
Hence we expect gauge bosons which couple to flavor-changing neutral
currents. In fact, this argument is slightly too slick: the same is true of
charged-current weak interactions as well!  However in that case the
gauge interactions, $SU(2)_W$ respect a global $(SU(5) \times U(1))^5$
chiral symmetry~\cite{ctsm} leading to the usual { GIM} mechanism.

Unfortunately, ETC interactions { cannot} respect GIM (exactly);
they must distinguish between the various generations in order to give
rise the masses of the different generations. Therefore, flavor-changing
neutral-current interactions are (at least at some level) unavoidable.

The most severe constraints come from possible $|\Delta S| = 2$
interactions which give rise to contributions to the $K_L$-$K_S$ mass
difference. In particular, we would expect that in order to produce
Cabbibo-mixing the same interactions which give rise to the $s$-quark
mass could give rise to the flavor-changing interaction
\beq
{\cal L}_{\vert \Delta S \vert = 2} = {g^2_{ETC} \, \theta^2_{sd} \over
{M^2_{ETC}}} \,\, \overline{s} \Gamma^\mu d \,\, \overline{s} \Gamma'_\mu d + {\rm
h.c.}~,
\eeq
where $\theta_{sd}$ is of order the Cabbibo angle. Such an interaction
contributes to the kaon mass splitting
\beq 
(\Delta M^2_K)_{ETC} =  {g^2_{ETC} \, \theta^2_{sd} \over
{M^2_{ETC}}} \, \langle \overline{K^0} \vert \overline{s} \Gamma^\mu d \, \overline{s}
\Gamma'_\mu d \vert K^0 \rangle + {\rm c.c.}
\eeq
Using vacuum insertion approximation we find
\beq
(\Delta M^2_K)_{ETC} \simeq {g^2_{ETC} \, 
{\rm Re}(\theta^2_{sd}) \over {2 M^2_{ETC}}} \,
f^2_K M^2_K ~.
\eeq
Experimentally we know that
$\Delta M_K <   3.5 \times 10^{-12}\,\mev $ and hence that
\beq
{M_{ETC} \over {g_{ETC} \, \sqrt{{\rm Re}(\theta^2_{sd})}}} >  600\,\tev
\eeq
Using eqn. \ref{fmass} we find that
\beq
m_{q, \ell} \simeq {g_{ETC}^2 \over {M_{ETC}^2}}
\langle\overline{T}T\rangle_{ETC}  < {0.5\,\mev\over{N_D^{3/2} \, \theta_{sd}^2}} \,
\eeq
showing that it will be difficult to get { $s$}-quark
mass right, let alone the { $c$}-quark!

\subsection{Pseudo-Goldstone Bosons}

As illustrated by our toy model above, a ``realistic'' ETC theory may
require a technicolor theory with a chiral symmetry structure bigger
than the $SU(2)_L \times SU(2)_R$ discussed in detail in the previous
lecture. The prototypical model of this sort is the one-family model
incorporated in our toy model. As discussed there the theory has an
$SU(8)_L \times SU(8)_R \to SU(8)_V$ $\Rightarrow$ chiral symmetry
breaking structure resulting in 63 Goldstone bosons, 60 of which
are physical. The quantum numbers of the 60 remaining Goldstone
bosons are shown in table \ref{pgbtab}. Clearly, these objects
cannot be massless in a realistic theory!

\begin{table}[htbp]
\begin{center}
\vskip.5pc
\setlength{\tabcolsep}{9pt}
\renewcommand{\arraystretch}{1.4}
\begin{tabular}{|c|c|c|}   \hline
SU$(3)_C$   & SU$(2)_{V}$ &Particle         \\ \hline
$1$      &$1$  &$P^{0 \prime} \;,\; \omega_T$  \\
$1$      &$3$  &$P^{0,\pm} \;,\; \rho^{0,\pm}_T$  \\
$3$      &$1$  &$P^{0 \prime}_{3} \;,\; \rho^{0 \prime }_{T 3}$  \\
$3$      &$3$  &$P^{0,\pm}_{3} \;,\; \rho^{0,\pm}_{T 3}$  \\
$8$      &$1$  &$P^{0 \prime}_{8} (\eta_T) \;,\; \rho^{0 \prime }_{T 8}$  \\
$8$      &$3$    &$P^{0,\pm}_{8} \;,\; \rho^{0,\pm}_{T 8}$  \\ \hline
\end{tabular}
\caption{Quantum numbers of the 60 physical Goldstone bosons (and the
corresponding vector mesons) in a one-family technicolor model. Note
that the mesons that transform as a 3's of QCD are complex fields.}
\end{center}
\label{pgbtab}
\end{table}

In fact, the ordinary gauge interactions break the full $SU(8)_L \times
SU(8)_R$ chiral symmetry explicitly. The largest effects are due to QCD
and the color octets and triplets mesons get masses of order 200 --- 300
GeV, in analogy to the electromagnetic mass splitting
$m_{\pi^+}-m_{\pi^0}$ in QCD. Unfortunately, the others~\cite{EL} are
{ massless} to {\cal O}($\alpha$)!

Luckily, the ETC interactions (which we introduced in order to 
give masses to the ordinary fermions) are capable of explicitly
breaking the unwanted chiral symmetries and producing masses for
these mesons. This is because in addition to coupling technifermions
to ordinary fermions, there will also be ETC interactions which couple
technifermions to themselves. Using Dashen's formula,
we can estimate that such an interaction can give rise to
an effect of order
\beq
F^2_{TC} M^2_{\pi_T} \propto {g^2_{ETC} \over M^2_{ETC}}
\langle (\overline{T}T)^2\rangle_{ETC}~.
\label{dashen}
\eeq
In the vacuum insertion approximation for a theory with small
$\gamma_m$, we may rewrite the above formula using eqn. \ref{fmass} and
find that
\beq
M_{\pi_T} \simeq 55\gev 
\sqrt{m_f \over 1\gev} \sqrt{250 \gev \over F_{TC}}~.
\eeq
It is unclear that this large enough.

In addition, there is a particularly troubling chiral symmetry
in the one-family model. The 
$SU(8)$-current $\overline{Q}\gamma_\mu \gamma_5 Q - 3 \overline{L} \gamma_\mu
\gamma_5 L$ is { spontaneously broken} and { has a color
  anomaly}. Therefore we have a potentially
dangerous { weak scale axion~\cite{axion}}!
An ETC-interaction of the form
\beq
{g^2_{ETC} \over M^2_{ETC}} \left(\overline{Q}_L \gamma^\mu L_L \right)
\left(\overline{L}_R \gamma^\mu Q_R \right)~,
\eeq
is required to give to an axion mass, therefore
{ we must~\cite{EL} embed $SU(3)_C$ in $ETC$.}

Finally, before moving on I would like to note that there is an implicit
assumption in the analysis of gauge-boson scattering presented in the
last lecture. We have assumed that {\it elastic} scattering dominates.
In the presence of many pseudo-Goldsone bosons, $WW$ scattering could
instead be dominated by {\it inelastic} scattering.  This effect has
been illustrated~\cite{hidden} in an $O(N)$-Higgs model with many
pseudo-Goldstone Bosons, solved in large-N limit.  Instead of the
expected resonance structure at high energies, the scattering can be
{ small and structureless} at all energies.

\subsection{ETC etc.\protect\cite{lanetasi}}

There are other model-building constraints on a realistic TC/ETC
theory. For completeness, I list them here:

\begin{itemize}

\item ETC should be asymptotically free.

\item There can be no gauge anomalies.

\item Neutrino masses, if nonzero, must be small.

\item There should be no extra massless, or light, gauge bosons.

\item Weak CP-violation, without strong CP-violation.

{ \item Isospin-violation in fermion masses without large $\Delta\rho$.}

{
\item Accomodate a large $m_t$.

\item Small corrections to $Z\to \overline{b}b$ and $b\to s\gamma$.}

\end{itemize}

Clearly, building a fully realistic ETC model will be quite difficult!
However, as I have emphasized before, this is because an ETC theory must
provide a complete dynamical explanation of flavor.  In the remainder of
this lecture, I will concentrate on possible solutions to the
flavor-changing neutral-current problem(s). As I will discuss in
detail in the last lecture, I believe the outstanding obstacle in
ETC or any theory of flavor is providing an explanation for the top-quark
mass, {\it i.e.} dealing with the last three issues listed
above.

\subsection{Technicolor with a Scalar}

At this point, it would be easy to believe that it is impossible
to construct a model of dynamical electroweak symmetry breaking.
Fortunately, there is at least an existence~\cite{ehs} proof of such
a theory: technicolor with a scalar.\footnote{Such a theory is
also the effective low-energy model for a ``strong-ETC'' theory
in which the ETC interactions {\it themselves} participate in
electroweak symmetry breaking.\cite{strongETC}} Admittedly, while electroweak
symmetry breaking has a dynamical origin in this theory, the introduction
of a scalar reintroduces  the hierarchy and naturalness problems we had
originally set out to solve.

\begin{figure}[tbp]
\hskip80pt\epsfxsize=6cm\epsfbox{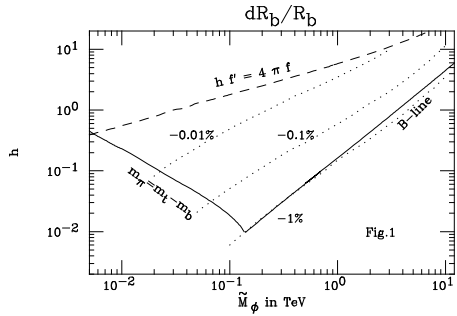}
\caption{Plot of allowed parameter space in model of technicolor
with a scalar.\protect\cite{carone} $h$ is the yukawa-coupling
of the scalar to technifermions and $\tilde{M}_\phi$ is the neutral scalar
mass. The ``triangular'' region formed by the solid and dashed
lines is allowed.}
\label{Fig10}
\end{figure}

In the simplest model one starts with a one doublet technicolor theory, 
and couples the chiral-symmetries of technifermions
to ordinary fermions through {\it scalar} exchange:
\beq
\hbox{\epsfxsize=2.5cm\epsfbox{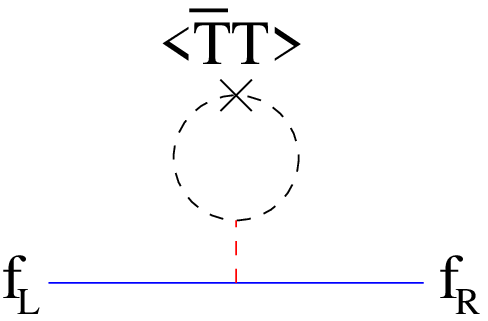}}
\eeq
The phenomenology of this model has been studied in detail,\cite{carone}
and the allowed region is shown in Figure \ref{Fig10}.

\subsection{Walking Technicolor and the Gap Equation}

Up to now we have assumed that technicolor is, like QCD, precociously
asymptotically free and $\gamma_m(\mu)$ is small for $\Lambda_{TC} < \mu
< M_{ETC}$. However, as discussed above it is difficult to construct an
ETC theory of this sort without producing dangerously large
flavor-changing neutral currents. On the other hand, if $\beta_{TC}$ is
{\it small}, $\alpha_{TC}$ can remain large above the scale
$\Lambda_{TC}$ --- {\it i.e.} the technicolor coupling would ``walk''
instead of run. In this same range of momenta, $\gamma_m$ may
be large and, since
\beq
\langle\overline{T} T\rangle_{ETC} = \langle\overline{T} T\rangle_{TC}
\exp\left(\int_{\Lambda_{TC}}^{M_{ETC}} {d\mu \over \mu}
\gamma_m(\mu)\right) 
\eeq
this could enhance $\langle \overline{T} T\rangle_{ETC}$
and fermion masses.\cite{walking}

\begin{figure}[tbp]
\hskip80pt\epsfxsize=6cm\epsfbox{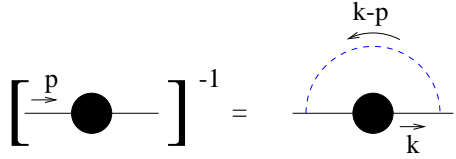}
\caption{Schwinger-Dyson equation for the
fermion self-energy function $\Sigma(p)$ in the rainbow
approximation.}
\label{Fig11}
\end{figure}

In order to proceed further, however, we need to understand how large
$\gamma_m$ can be and how walking affects the technicolor $\chi$-symmetry
breaking dynamics.  These questions cannot be addressed in perturbation
theory.  Instead, what is conventionally done is to use a 
nonperturbative aproximation for $\gamma_m$ and chiral-symmetry breaking
dynamics based on the ``rainbow'' approximation~\cite{gap}
to Schwinger-Dyson equation for shown in Figure \ref{Fig11}.
Here we write the full, nonperturbative, 
fermion propagator in momentum space as
\beq
iS^{-1}(p) = Z(p)(\slashchar{p} - \Sigma(p))~.
\eeq

The linearized form of the gap equation in
Landau gauge (in which $Z(p) \equiv 1$ in the rainbow
approximation) is
\beq
\Sigma(p) = 3 C_2(R)\, \int {d^4 k \over {(2 \pi)^4}}
\, {\alpha_{TC}((k-p)^2) \over {(k-p)^2}} \, {\Sigma(k) \over {k^2}}~.
\eeq
Being separable, this integral equation can be converted to a
differential equation which has the approximate (WKB)
solutions~\cite{wkb} (here $\alpha(\mu)$ is assumed to run slowly, as
will be the case in walking technicolor):
\beq
\Sigma(p) \propto p^{-\gamma_m(\mu)}\, ,\, \, p^{\gamma_m(\mu)-2}
\eeq
where the anomalous dimension of the fermion mass operator is
\beq
\gamma_m(\mu)=1-\sqrt{1-{\alpha_{TC}(\mu)\over\alpha_C}}\, ; \, \, \alpha_C
\equiv {\pi \over 3 C_2(R)}\, .
\label{crit}
\eeq

One can give a physical interpretation of these two
solutions.\cite{politzer} Using the operator product expansion, we find
\beq
\lim_{p\to\infty} \Sigma(p) \propto
\lower15pt\hbox{\epsfxsize=4cm\epsfbox{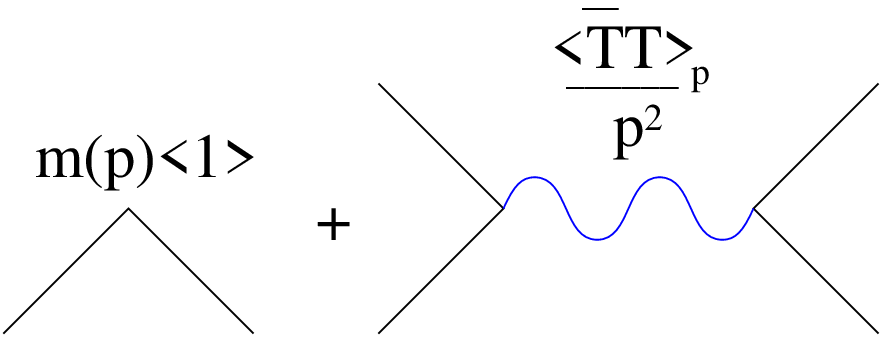}}~,
\eeq
and hence the first solution corresponds to a ``hard mass'' or explicit
chiral symmetry breaking, while the second solution corresponds to a
``soft mass'' or spontaneous chiral symmetry breaking.  If we let $m_0$
be the explicit mass of a fermion, dynamical symmetry breaking occurs
only if
\beq
\lim_{m_0 \to 0} \Sigma(p) \neq 0\, .
\eeq
A careful analysis of the gap equation, or equivalently the appropriate
effective potential,\cite{cjt} implies that this happens only if
$\alpha_{TC}$ reaches a critical value of chiral
symmetry breaking, $\alpha_C$ defined in eqn. \ref{crit}.
Furthermore, the chiral symmetry breaking scale $\Lambda_{TC}$ is
defined by the scale at which
\beq
\alpha_{TC}(\Lambda_{TC})=\alpha_C 
\eeq
and hence, at least in the rainbow approximation, at which
\beq
\gamma_m(\Lambda_{TC})=1.
\eeq
In the rainbow approximation, then, chiral symmetry breaking occurs when
the { ``hard''} and { ``soft''} masses scale the same way. It
is believed that even beyond the rainbow
approximation $\gamma_m =1$ at the critical coupling.\cite{cohen}

\subsection{Implications of Walking: Fermion and PGB Masses, $S$}

If $\beta(\alpha_{TC}) \simeq 0$ all the way from $\Lambda_{TC}$
to $M_{ETC}$, then  $\Rightarrow$ $\gamma_m(\mu) \cong 1$ in this
range. In this case, eqn. \ref{fmass} becomes
\beq
m_{q,l} = {g^2_{ETC} \over {M^2_{ETC}}} \times
\left(\langle\overline{T}T\rangle_{ETC} \cong 
\langle\overline{T}T\rangle_{TC} \, {M_{ETC} \over {\Lambda_{TC}}} \right)~.
\eeq
We have previously estimated that flavor-changing
neutral current requirements imply that the
ETC scale associated with the second generation must
be greater than of order 100 to 1000 TeV. In the case of walking
the enhancement of the technifermion condensate implies that
\beq
m_{q,l} \simeq {{50\, -\, 500\mev}\over N^{3/2}_D \theta^2_{sd}}~,
\eeq
arguably enough to accomodate the strange and charm quarks.

While this is very encouraging, two { caveats} should be kept in
mind. First, the estimates given are for limit of { ``extreme
  walking''}, {\it i.e.} assuming that the technicolor coupling walks
all the way from the technicolor scale $\Lambda_{TC}$ to the relevant
ETC scale $M_{ETC}$. To produce a more complete analysis, ETC-exchange
must be incorporated into the gap-equation technology in order to
estimate ordinary fermion masses. Studies of this sort are encouraging,
it appears possible to accomodate the first and second generation masses
without necessarily having dangerously large flavor-changing
neutral currents.\cite{walking} The second issue, however, is
what about the third generation quarks, the { top} and { bottom}?
As we will see in the next lecture, because of the large top-quark
mass, further refinements or modifications will be necessary to produce
a viable theory of dynamical electroweak symmetry breaking.

In addition to modifying our estimate of the relationship
between the ETC-scale and ordinary fermion masses, walking
also influences the size of pseudo-Goldstone boson masses. 
In the case of walking, Dashen's formula for the
size of pseudo-Goldstone boson masses in the presence
of chiral symmetry breaking from ETC interactions, eqn. \ref{dashen},
reads:
\beqa
F^2_{TC} M^2_{\pi_T} & \propto & {g^2_{ETC} \over M^2_{ETC}}
\langle \left(\overline{T}T\right)^2\rangle)_{ETC} \nonumber \\
&\approx& {g^2_{ETC} \over M^2_{ETC}} 
\left(\langle\overline{T}T\rangle_{ETC}\right)^2 \nonumber \\
&\simeq&{g^2_{ETC} \over M^2_{ETC}}
{M^2_{ETC}\over \Lambda^2_{TC}}
\left(\langle\overline{T}T\rangle_{TC}\right)^2~,
\eeqa
where, consistent with the rainbow approximation, we have used the vacuum-insertion 
to estimate the strong matrix element.
Therefore we find
\beqa
M_{\pi_T} & \simeq &  g_{ETC} 
\left({4\pi F^2_{TC} \over \Lambda_{TC}}\right) \nonumber\\
&\simeq &g_{ETC} \left({750\gev \over N_D}\right)
\left({1\tev\over \Lambda_{TC}}\right)~,
\eeqa
{\it i.e.} walking also enhances the size of pseudo-Goldstone
boson mases!

Finally, { {what about S?}} As emphasized by Lane,\cite{lanetasi} the
assumptions of previous estimate of $S$ included~\cite{oblique} that:
\begin{itemize}

{ 
\item techni-isospin is a good symmetry, and

\item { Technicolor is QCD-like, {\it i.e.}.}}
\begin{enumerate}

\item Weinberg's sum rules are valid,

\item the spectral functions saturated by lowest-resonances,

\item that the masses/couplings of resonances can be scaled from QCD.

\end{enumerate}

\end{itemize}
A ``realistic'' walking technicolor theory would be very unlike QCD:
\begin{itemize}

{
\item Walking $\Rightarrow$ different behavior of spectral functions.

\item Many flavors/PGBs and non-fundamental representations makes
  scaling from QCD suspect.  
}
\end{itemize}
For this reason the analysis given previously does not apply, and a
walking theory could be phenomenologically acceptable.  Unfortunately,
technicolor being a strongly-coupled theory, it is not possible to give
a compelling argument that the value of $S$ in a walking technicolor
theory {\it is} definitely acceptable.

\section{Lecture 3: Top in Models of Dynamical Symmetry Breaking}

\subsection{The ETC of $m_t$}

Because of it's large mass, the top quark poses a particular
problem in models of dynamical electroweak symmetry breaking.
Consider an ETC interaction ({\it c.f.} eqn. \ref{fmass})
giving rise to the top quark mass
\beq
{\lower15pt\hbox{\epsfysize=0.5 truein \epsfbox{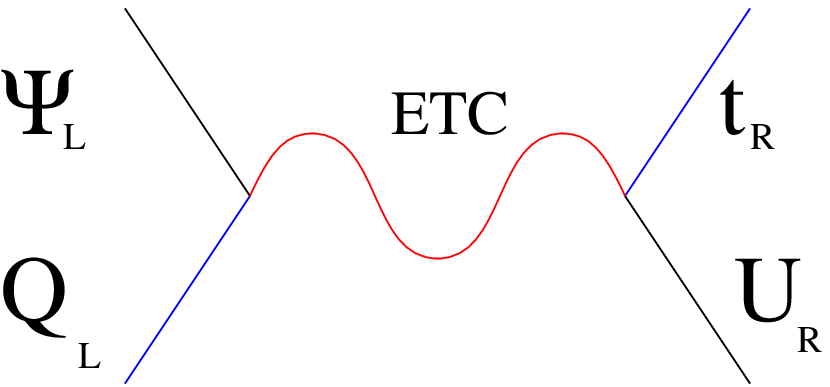}}}
\Rightarrow {{g_{ETC}^2\over M^2_{ETC}}}(\overline{\Psi}_L U_R)
({\overline{t}_R Q_L})~,
\eeq
yielding
\beq
m_t \approx {{g_{ETC}^2\over M^2_{ETC}}} \langle\overline{U} U\rangle_{ETC}~.
\eeq
In conventional technicolor, using
\beq
\langle\overline{U} U\rangle_{ETC} \approx \langle\overline{U} U\rangle_{TC}
\approx 4\pi F^3_{TC}
\eeq
we find
\beq
{ {M_{ETC}\over g_{ETC}}} \approx 1 \tev 
\left({F_{TC}\over 250\gev}\right)^{3\over 2}
\left({175 \gev \over m_t}\right)^{1\over 2}~.
\label{tmass}
\eeq
That is, { the scale of top-quark ETC-dynamics is {\it very} low.}
Since $M_{ETC} \simeq \Lambda_{TC}$ and
\beq
\langle\overline{U} U\rangle_{ETC} = \langle\overline{U} U\rangle_{TC}
\exp\left(\int_{\Lambda_{TC}}^{M_{ETC}} {d\mu \over \mu}
\gamma_m(\mu)\right)~,
\eeq
we see that { walking} can't alter this conclusion.\cite{walkingzbb}
As we will see in the next few sections, a low ETC scale for the
top quark is very problematic.

\subsection{ETC Effects~\protect\cite{zbb,walkingzbb} on $Z \rightarrow b \overline{b}$}

For ETC models of the sort discussed in the last lecture, in which the
ETC gauge-bosons do not carry weak charge, the gauge-boson responsible
for the top-quark mass couples to the current
\beq                    
\xi (\overline{\Psi}^{i\alpha}_L \gamma^\mu { Q^i_L}) +
\xi^{-1} (\overline{U}^\alpha_R \gamma^\mu { t_R})~,
\label{current}
\eeq                                                   
(or $h.c.$) where $\alpha$ is the technicolor index and and the
contracted $i$ are weak-indices.  The part of the exchange-interaction
coupling left- and right-handed fermions leads to the top-quark mass.

Additional interactions arise from the same dynamics, including
\beq
-{ {g^2_{ETC}\over M^2_{ETC}}}
({ \overline{t}_R}\gamma^\mu U^\alpha_R)
(\overline{U}^\alpha_R\gamma_\mu { t_R})
\eeq
and
\beq
-{ {g^2_{ETC}\over M^2_{ETC}}}
({ \overline{Q}^i_L}\gamma^\mu \Psi^{i\alpha}_L)
(\overline{\Psi}^{j\alpha}_L\gamma_\mu { Q^j_L})~.
\eeq
The last interaction involves { $b_L$} and the { technifermions}.
After a Fierz transformation, the left-handed operator becomes the
product of weak triplet currents
\beq
-{1\over 2}{ {g^2_{ETC}\over M^2_{ETC}}}
({ \overline{Q}^i_L \gamma^\mu \tau_a^{ij} Q^j_L})
(\overline{\Psi}^k_L \gamma_\mu \tau_a^{kl} \Psi^l_L)~,
\eeq
where the $\tau$ are the Pauli matrices, plus terms involving
weak singlet currents (which will not concern us here).

The exchange of this ETC gauge-bosons produces
a correction of the coupling of the $Z$ to $b\bar{b}$
\beq
\hskip-10pt{\lower35pt\hbox{
\epsfysize=0.75 truein \epsfbox{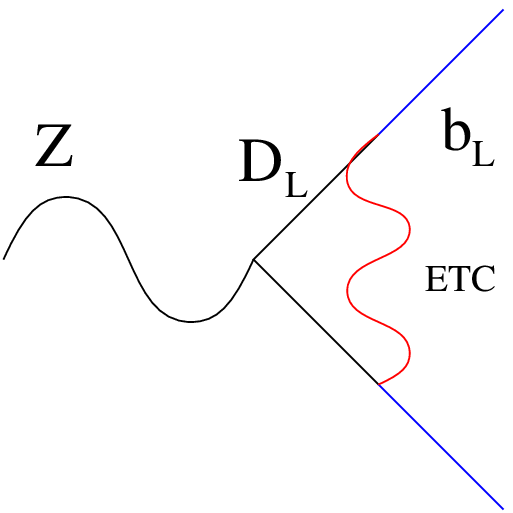}}
\hskip10pt
\rightarrow  
\hskip10pt
{\lower25pt\hbox{
\epsfysize=0.75 truein \epsfbox{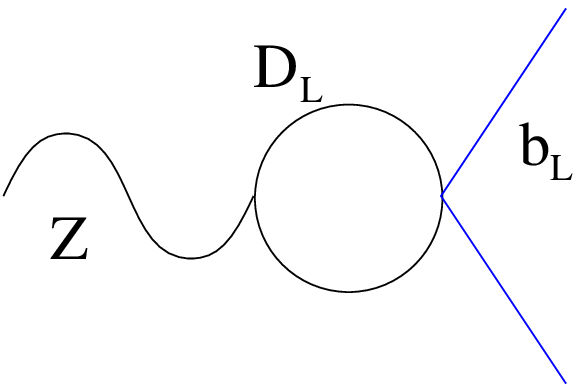}}}
}~.
\eeq
The size of this effect can be calculated by comparing it
to the technifermion weak vacuum-polarization diagrm
\beq
\hskip-10pt{\lower7pt\hbox{
\epsfxsize=1.0 truein \epsfbox{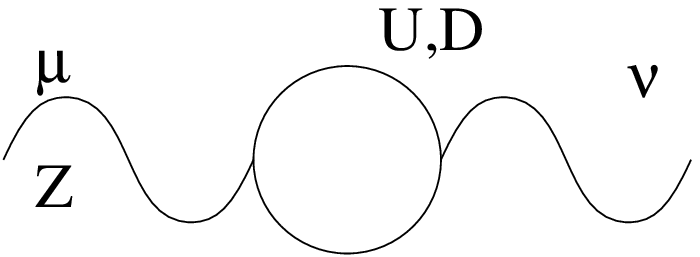}}
\rightarrow  \pi^{\mu\nu}_{ij} = 
\left(q^2 g^{\mu\nu} - q^\mu q^\nu\right) \delta_{ij} \pi(q^2)\, ,}
\label{zbbetc}
\eeq
which, by the Higgs mechanism yields
\beq
\pi(q^2) = \frac{e^2 v^2}{4 \sin^2_\theta \cos^2\theta} {1\over q^2}\, .
\eeq

Therefore, exchange of the ETC gauge-boson responsible
for the top-quark mass leads to a low-energy effect which
can be summarized by the operator
\beq
-{e\over 2 \sin\theta \cos\theta} { g^2_{ETC} v^2\over M^2_{ETC}}
\xi^2 (\overline{Q}_L \slashchar{Z} \tau_3 Q_L)~.
\eeq
Hence this effect results
in a change in the $Zb\overline{b}$ coupling 
\beq
\delta g_L = +{1\over 4}
{e \over \sin\theta \cos\theta} 
\, \xi^2 {g^2_{ETC} v^2 \over M^2_{ETC}}\, ,
\eeq
which, using the relation in eqn. \ref{tmass}, 
results in
\beq
{\delta \Gamma_b \over \Gamma_b} \approx 
{2 g_L \delta g_L \over  g^2_L + g^2_R} \approx
-6.5\% \cdot \xi^2 \cdot \left({m_t \over 175 \gev}\right)\, .
\label{dgamma}
\eeq
It is convenient to form the ratio $R_b = \Gamma_b/\Gamma_h$, where
$\Gamma_b$ and $\Gamma_h$ are the width of the $Z$ boson to $b$-quarks
and to all hadrons, respectively, since this ratio is largely
independent of the ``oblique'' corrections $S$ and $T$. The
shift in eqn. \ref{dgamma} results in a shift in $R_b$ of
approximately
\beq
{ {\delta R_b \over R_b }} \approx
{\delta \Gamma_b \over \Gamma_b}(1-R_b) \approx 
{ -5.1\%} \cdot \xi^2 \cdot \left({m_t \over 175 \gev}\right)\, .
\eeq

\begin{figure}[tbp]
\hskip80pt
\epsfxsize=6cm
\centerline{\epsffile{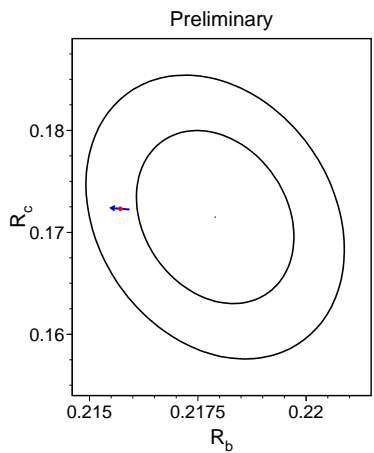}}
\caption{Contours in the $R_b$-$R_c$ plane from LEP data,\protect\cite{lepewwg}
corresponding to 68\% and 95\% confidence levels assuming Gaussian systematic
errors. The Standard Model prediction for $m_t$=175$\pm$6 GeV is
also shown. The arrow points in the direction of increasing values of 
$m_t$.}
\label{Fig12}
\end{figure}

Recent LEP results~\cite{lepewwg} on $R_b$ are shown in Figure \ref{Fig12}.
As we see, the current experimental value of $R_b$ is about 1.8$\sigma$
{\it above} the standard model prediction, while a shift of -5.1\% would
(given the current experimental plus systematic experimental
error~\cite{lepewwg} one $\sigma$ corresponds to a shift of 0.7\%) {\it
  lower} $R_b$ by approximately 7$\sigma$! Clearly, conventional ETC
generation of the top-quark mass is ruled out.

It should be noted, however, that there are nonconventional ETC models
in which $R_b$ {\it may not} be a problem. The analysis leading to the
result given above assumes that (see eqn. \ref{current}) the ETC
gauge-boson responsible for the top-quark mass {\it does not} carry
weak-$SU(2)$ charge.  It is possible to construct models~\cite{noncom}
where this is not the case.  Schematically, the group-theoretic
structure of such a model would be as follows
\medskip
\begin{center}

$ETC  \times SU(2)_{light}$

$\ \ \ \ \ \downarrow\ \ \ \ \ f $

$TC \times SU(2)_{heavy}  \times SU(2)_{light}$

$\ \ \ \ \ \downarrow\ \ \ \ \ u $

$TC  \times SU(2)_{weak} $

\end{center}
\medskip\noindent
where ETC is extended technicolor, $SU(2)_{light}$ is (essentially)
weak-$SU(2)$ on the light fermions, $SU(2)_{heavy}$ (originally embedded
in the ETC group) is weak-$SU(2)$ for the heavy fermions, and
$SU(2)_{light} \times SU(2)_{heavy}$ break to their diagonal subgroup
(the conventional weak-interactions, $SU(2)_{weak}$) at scale $u$.

In this case a {\it weak-doublet, technicolored} ETC boson 
coupling to
\beq
\xi\overline{Q}_L \gamma^\mu U_L + {1\over \xi}\bar t_R \gamma^\mu \Psi_R~,
\eeq
is responsible for producing $m_t$.
A calculation analogous to the one above yields a correction
\beq
\hskip-20pt{\lower25pt\hbox{
\epsfysize=0.75 truein \epsfbox{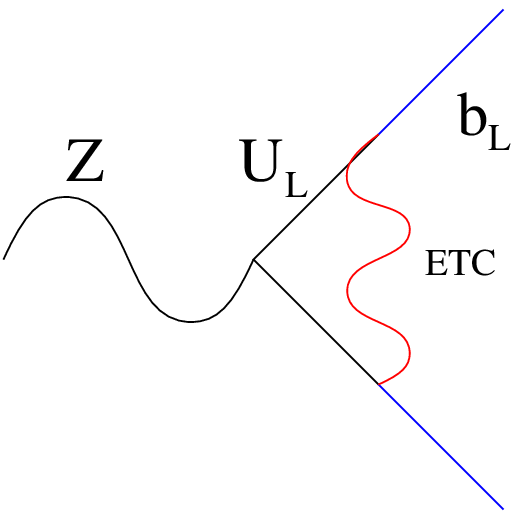}}
\rightarrow  
\delta g_L = -{1\over 4}
{e \over \sin\theta \cos\theta} 
\, \xi^2 {g^2_{ETC} v^2 \over M^2_{ETC}}
}
\eeq
of the {\it opposite} sign. In fact, the situation is
slightly more complicated: there is also an extra
{  $Z$-boson} which also contributes. The 
total contribution is found~\cite{noncom} to be
\beq
{\delta R_b \over R_b} \approx +5.1\% \cdot \xi^2 \cdot 
\left({m_t\over 175 {\rm GeV}}\right)\left( 1 - 
{\sin^2 \alpha \over \xi^2} {f^2\over u^2}\right)
\eeq
where $\tan\alpha=g^\prime/g$ is the ratio of the $SU(2)_{light}$
and $SU(2)_{heavy}$ coupling constants. The overall contribution
to $R_b$ is very model-dependent, but could be within the
experimentally allowed window.

\subsection{Isospin Violation: $\Delta\rho$}

\medskip
\noindent\underline{``Direct'' Contributions}
\medskip

ETC-interactions {\it must} violate weak-isospin in order to give rise
to the mass splitting between the top and bottom quarks. This could
induce dangerous $\Delta I=2$ { technifermion}
operators~\cite{appelquistrho}
\beq
{\lower10pt\hbox{\epsfxsize=1.5in \epsffile{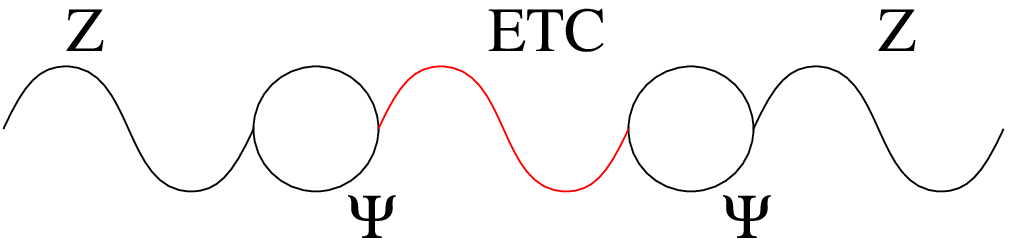}}}
\Rightarrow { g^2_{ETC}\over M^2_{ETC}}
\left(\overline{\Psi}_R \gamma_\mu \tau_3 \Psi_R\right)^2\, .
\eeq
We can estimate the contribution of these operators to $\Delta\rho$
using the vacuum-insertion approximation
\beq
\Delta\rho   \simeq  {2 g^2_{ETC}\over M^2_{ETC}}
{N^2_D F^4_{TC} \over v^2} 
\eeq
which yields
\beq
\Delta\rho \approx  12\% \cdot 
\left({\sqrt{N_D} F_{TC} \over 250 \gev}\right)^2
\cdot \left({1 \tev \over M_{ETC}/g_{ETC}}\right)^2\, .
\eeq
If we require that $\Delta \rho \le 0.4\%$, we find
\beq
{M_{ETC} \over g_{ETC}} > 5.5\tev \cdot
\left({\sqrt{N_D} F_{TC} \over 250 \gev}\right)^2\, ,
\eeq
{\it i.e.} $M_{ETC}$ must be { greater} 
than required for $m_t \simeq 175 \gev$.

There is { another possibility}.  It is possible that $N_D F^2_{TC}
\ll (250 \gev)^2$, { if} the sector responsible for the top-quark
mass { does not} give rise to the bulk of EWSB.  In this scenario,
the constraint is
\beq
F_{TC}< {105 \gev \over N^{1/2}_D} \cdot
\left({M_{ETC}/g_{ETC} \over 1 \tev}\right)^{1/2}\, .
\eeq
However, this modification would {\it enhance} the effect of
ETC-exchange in $Z \to b \overline{b}$.

\medskip
\underline{``Indirect'' Contributions} to $\Delta\rho$
\medskip

\begin{figure}[tbp]
\hskip80pt
\epsfxsize=6cm
\epsffile{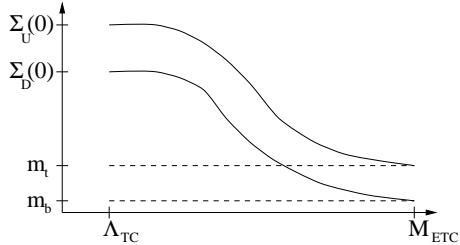}
\caption{Momentum dependent dynamical masses of the technifermions
responsible for the $t$- and $b$-quark masses, based on an
a gap-equation analysis.}
\label{Fig13}
\end{figure}

Isospin violation in the ordinary fermion masses suggests
the existence of isospin violation in the technifermion dynamical
masses. Indeed, an analysis of the gap equation shows that
if the $t$- and $b$-quarks get masses from technifermions in
the same technidoublet the dynamical masses of the corresponding
technifermions are as shown in Figure \ref{Fig13}. At a scale
of order $M_{ETC}$ the technifermions and ordinary fermions are
unified into a single gauge group, so it is not surprising
that their masses are approximately equal at that scale. Below
the ETC scale, the technifermion dynamical mass runs (because of the
technicolor interactions), while the ordinary fermion masses
do not. As shown in Figure \ref{Fig13}, therefore,
we expect that $\Sigma_U(0) - \Sigma_D(0)$
\gae$m_t - m_b$.

We can estimate the contribution of this effect to $\Delta\rho$
\beq
\lower10pt\hbox{\epsfxsize=1.0in\epsffile{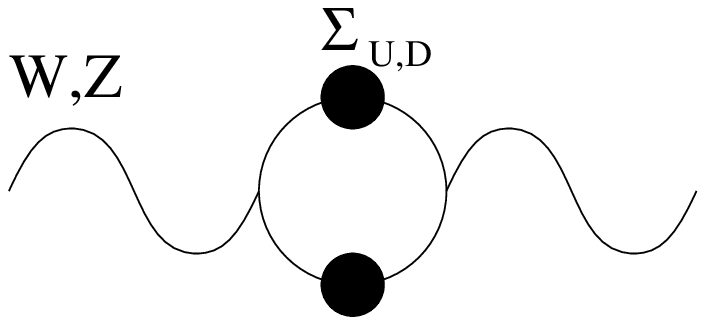}}
\propto {N_D d \over 16\pi^2}{(\Sigma_U(0) - \Sigma_D(0))^2\over v^2}\, ,
\eeq
where $N_D$ is the number of technidoublets 
and $d$=dimension of TC representation. If we
require $\Delta\rho \le 0.4\%$, this yields
\beq
N_D d  \left({\Delta\Sigma(0)\over 175 \gev}\right)^2
\le 2.7~.
\eeq
This is perhaps possible if $N_D=1$ and $d=2$ ({\it i.e.}
$N_{TC}=2$), but is generally problematic.

\subsection{Evading the Unavoidable}

\begin{figure}[tbp]
\hskip80pt
\epsfxsize=6cm
\epsffile{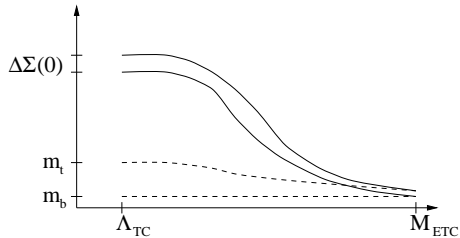}
\caption{Momentum dependent dynamical masses of the technifermions
  which couple to the $t$- and $b$-quarks in a theory where additional
  strong interactions (other than techicolor) are responsible for the
  bulk of the top and bottom quark masses. These additional strong
  interactions allow for the quark masses to run significantly below the
  ETC scale.}
\label{Fig14}
\end{figure}

The problems outlined in the last two sections, namely potentially
dangerous ETC corrections to the branching ratio of $Z\to b\overline{b}$
and to the $\rho$ parameter, rule out the possibility of generating the
top-quark mass using conventional extended technicolor interactions.  A
close analysis of these problems, however, suggests a framework for
constructing an acceptable model: arrange for the $t$- and $b$-quarks to
get the majority of their masses from interactions { other than
  technicolor}. If this is the case, the top- and bottom-quark
masses can run substantially below the ETC scale as shown in
Figure \ref{Fig14}, allowing for 
\beq
\hskip-10pt{\Delta \Sigma(0) \simeq m_t(M_{ETC})-m_b(M_{ETC}) \ll m_t}~.
\eeq
Since the technicolor/ETC interactions would only be responsible for a
{\it portion} of the top-quark mass in this type of model, the problems
outlined in the previous two sections are no longer relevant.  In order
to produce a substantial running of the third-generation quark masses,
the third-generation fermions must have an additional {
  strong-interaction} not shared by the first two generations of
fermions or (at least in an isospin-violating way) by the
technifermions.

\subsection{An Aside: Top-Condensate Models}

Before constructing a model of the sort proposed in last section, we
should pause to consider another possibility.  Having entertained the
notion that the top-quark mass may come from a strong interaction felt
(at least primarily) by the third generation, one should ask if there is
any longer a need for technicolor! After all, any interaction that gives
rise to a quark mass {\it must} break the weak interactions.
Furthermore, since $m_t \simeq M_W,\, M_Z$, the top-quark is much
heavier than other fermions it must be more more strongly coupled to
symmetry-breaking sector.  Perhaps all~\cite{topmode} of
electroweak-symmetry breaking is due to a condensate of top-quarks,
$\langle \bar{t}t\rangle \neq 0$.

\begin{figure}[tbp]
\hskip80pt
\epsfxsize=6cm
\epsffile{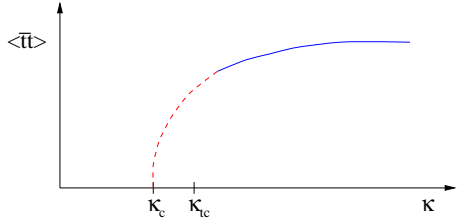}
\caption{Behavior of the condensate in a top-color
model as a function of the top-color coupling assuming
a {\it continuous} transition.}
\label{Fig15}
\end{figure}

Consider a { spontaneously broken/strong gauge-interaction},
e.g. top-color:
\beq
{ SU(3)_{tc}} \times { SU(3)}
\stackrel{M}{\to} SU(3)_{QCD}~,
\eeq
where $SU(3)_{tc}$ is a new, strong, top-color interaction coupling
to the third-generation quarks and the other $SU(3)$ is a weak, color
interaction coupling to the first two generations. At scales below
$M$, one has the ordinary QCD plus interactions which
couple primarily to the third generation quarks 
and can be summarized by an operator of the form
\beq
{\cal L} \supset - {4\pi\kappa \over M^2}
\left(\overline{Q}\gamma_\mu {\lambda^a\over 2} Q\right)^2~,
\eeq
where $\kappa \approx g^2_{tc}/4\pi$ is related to the top-color
coupling constant. Consider what happens as, for fixed $M$, we
vary $\kappa$. For small $\kappa$, the interactions are perturbative
and there is no chiral symmetry breaking. For large $\kappa$, since the
new interactions are attractive in the spin-zero, isospin-zero
channel, we expect chiral symmetry breaking with
$\langle \bar{t}t\rangle \propto M^3$. If the transition between these
two regimes is {\it continuous}, as it is in the bubble~\cite{NJL}
or mean-field approximation, we expect that the condensate
will behave as shown in Figure \ref{Fig15}.

In order to produce a realistic model of electroweak symmetry breaking
based on these considerations, one must introduce extra interactions to
split the top- and bottom-quark masses.  A careful analysis then shows
that it is not possible to achieve a phenomenologically acceptable
theory unless~\cite{topmode} the scale $M \gg v$. Since the weak
scale $v$ is fixed, this implies that the condensate 
$\langle \bar{t}t\rangle \ll M^3$, and the top-color
coupling $\kappa$ must be finely tuned
\beq
{\Delta\kappa \over \kappa_c} \equiv
{{\kappa-\kappa_c}\over \kappa_c}
\propto {\langle\bar{t}t\rangle \over M^3}~.
\eeq
In this region, one has simply reproduced the
standard-model,\cite{topmode} with the Higgs-boson $\phi$ produced
dynamically as a $\overline{t_R}Q_L$ bound state!

\subsection{Topcolor-Assisted~\protect\cite{tcii} Technicolor (TC2)}

Recently, Chris Hill has proposed~\cite{tcii} a theory which combines
technicolor and top-condensation. Features of this type of model include
\begin{itemize}

\item Strong Technicolor dynamics at 1 TeV which dynamically generates
{ most} of electroweak symmetry breaking;

\item Extended Technicolor dynamics at scales much higher than 1 TeV
which generate the light quark and lepton masses, { and small contributions
to the third generation masses} ($m_{t,b,\tau}^{ETC}$) of order 1 GeV;

\item Strong Topcolor dynamics { also} at a scale of order 1 TeV
which generates $\langle \bar{t} t \rangle \neq 0$, $m_t \sim 175$ GeV;

\item Topcolor {\bf does not} form $\langle \bar{b} b \rangle$,
and therefore there must be isospin violation. This may be
acceptable because...

\item Topcolor contributes a small amount to EWSB (with an
  ``F-constant'' $f_t \sim 60$ GeV);

\item Extra pseudo-Goldstone bosons (``Top-pions'') which get mass from
  ETC interactions which allow for mixing of third generation to first
  two.

\end{itemize}
\medskip
{\underline{Hill's Simplest TC2 Scheme}}
\medskip

The simplest scheme~\cite{tcii} which realizes these features has the
following structure:
\medskip
\begin{center}
$G_{TC}  \times  SU(2)_{EW} \times $
\end{center}
\begin{center}
$SU(3)_{tc} \times SU(3) \times U(1)_H \times U(1)_L$
\end{center}
\begin{center}
${ (g^{tc}_3 > g_3)}\ \ \downarrow \ \ M$ \gae 1 TeV$\ \ { (g^H_1 > g^L_1)}$
\end{center}
\begin{center}
$G_{TC} \times SU(3)_C  \times SU(2)_{EW} \times U(1)_Y $
\end{center}
\begin{center}
$\downarrow\ \ \ \Lambda_{TC}\sim 1{\ \rm TeV} $
\end{center}
\begin{center}
$SU(3)_C \times U(1)_{EM}$
\end{center}
\medskip
Here $U(1)_H$ and $U(1)_L$ are $U(1)$ gauge groups coupled to the
(standard model) hypercharges of the third-generation and first-two generation
fermions respectively.
Below $M$, this leads to the effective interactions:
\beq
-{{4\pi \kappa_{tc}}\over{M^2}}\left[\overline{\psi}\gamma_\mu 
{{\lambda^a}\over{2}} \psi \right]^2~,
\eeq
from top-color exchange and the isospin-violating interactions
\beq
-{{4\pi \kappa_1}\over{M^2}}\left[{1\over3}\overline{\psi_L}\gamma_\mu  \psi_L
+{4\over3}\overline{t_R}\gamma_\mu  t_R
-{2\over3}\overline{b_R}\gamma_\mu  b_R
\right]^2\, ,
\label{hyper}
\eeq
from exchange of the ``heavy-hypercharge'' ($Z^\prime$) 
gauge boson.

Since the interactions in eqn. \ref{hyper} are attractive in the
$\bar{t}t$ channel, but repulsive in the $\bar{b}b$ channel, the
couplings $\kappa_{tc}$ and $\kappa_1$ can be chosen to produce $\langle
\bar{t} t \rangle \neq 0$ and a large $m_t$, but {\bf not} $\langle
\bar{b} b \rangle = 0$. In the Nambu-Jona-Lasinio approximation,\cite{NJL}
we require
\beq
 \kappa^t = \kappa_{tc} +{1\over3}\kappa_1 >
\kappa_c \left( = {{3\pi}\over{8}} \right)_{NJL} >
\kappa^b=\kappa_{tc} -{1\over 6}\kappa_1~ .
\label{fine}
\eeq

\subsection{$\Delta \rho$ in TC2~\protect\cite{terning}}
\medskip
\underline{Direct Contributions}
\medskip

Couplings of the (potentially strong) $U(1)_H$ group are isospin
violating, at least in it's couplings to the third generation.  Isospin
violating couplings to technifermions could be very
dangerous,\cite{terning} as shown above.  For example, in the one-family
technicolor model, if $U(1)_H$ charges proportional to $Y$:
\beq
\Delta \rho^{\rm T} \approx 152\% \
\kappa_1 \left({{1\ {\rm TeV}}\over{M}}\right)^2~.
\eeq
If $M\simeq 1\tev$, we must have $\kappa_1 \ll 1$. From eqn. \ref{fine}
above, this implies a { fine tuning} of $\kappa_{tc}$.  In order to
avoid this problem, one must construct a model in which the {
  $U(1)_H$ couplings to technifermions are isospin
  symmetric~\cite{natural} -- ``Natural TC2''}.

\medskip
\underline{Indirect/Direct Contribution}
\medskip

Since there are additional (strong) interactions felt by the
third-generation of quarks, there are new 
``two-loop'' contributions~\cite{terning} to $\Delta \rho$:
\beq
\lower7pt\hbox{\epsfxsize 3cm {\epsffile{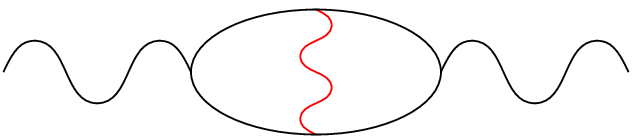}}}
\Leftrightarrow
\lower7pt\hbox{\epsfxsize 3cm {\epsffile{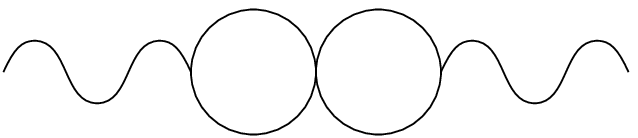}}}~.
\eeq
This contribution yields
\beq
\hskip-15pt{\Delta \rho^{\rm tc} \approx 0.53\%
\left({\kappa_{tc}\over \kappa_c}\right)
\left(1\ {\rm TeV}\over M\right)^2
\left(f_t \over 64\ {\rm GeV}\right)^4\, .}
\eeq
From this we find that $M$ \gae 1.4 TeV.

\subsection{Electroweak Constraints~\protect\cite{constraints} on Natural TC2}

\begin{figure}[tbp]
\hskip80pt
\epsfxsize=6cm
\centerline{\epsfbox{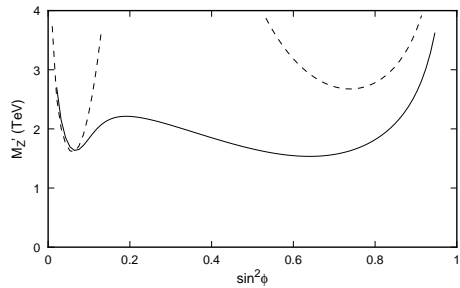}}
\caption{Bounds~\protect\cite{constraints}
on the mass of the $Z^\prime$ in natural
TC2 models as a function of the angle $\phi$ 
where $\tan\phi = g^L_1/g^H_1$.
Bounds are shown for $\alpha_s(M_Z)=0.115$ (solid), 0.124 (dashed).}
\label{Fig16}
\end{figure}

If the $U(1)_H$ couplings to technifermions are isospin-symmetric,
electroweak phenomenology is specified by $M^2_{Z^\prime}$, $\tan\phi =
g^L_1/g^H_1$, and the charges $Y_H$ of ordinary fermions. To get a
feeling for the size of constraints on these models from electroweak
phenomenology, consider a ``baseline'' model: $Y_H = Y$.  While this may
be unrealisic, it is flavor universal. In this case the third generation
picked out by it's couplings to $SU(3)_H$.

Constraints (arising from $Z$-$Z^\prime$ mixing as well
as $Z^\prime$ exchange) from all precision electroweak data
are shown in Figure \ref{Fig16}. We see that, even in light of
current LEP data, natural TC2 with a $Z^\prime$ mass of order
1-2 TeV is allowed.

\section{Where have we come from, where are we going?}

In these lectures I have tried to provide an introduction to modern
theories of dynamical electroweak symmetry breaking. We have come a long
way, and it is worth reviewing the logical progression that has brought
us here: 
\medskip
\begin{itemize}

\item The search for a {\bf natural} and {\bf dynamical} explanation for
  electroweak symmetry breaking implies we should explore technicolor
  and related models (lecture 1);

\item Accommodating/explaining $u,\, d,\, s,\, c$ masses in such
  theories without large {\bf flavor-changing neutral-currents}
  leads us to consider ``walking'' technicolor (lecture 2);

\item Accommodating the $b$ and, especially, the $t$ mass without large
  corrections to { $\Delta \Gamma_b$} and {$T$} leads us to consider
  top-color assisted technicolor related model(s) (lecture 3).
\end{itemize}
\medskip

Despite the progress that has been made, no complete and
consistent model exists. As I have emphasized, model building
difficult becuase
\medskip
\begin{itemize}

\item Technicolor is a non-decoupling theory, the natural dynamical
scale must be of order 1 TeV. Therefore, there are always potentially
large low-energy effects;

\item Technicolor theories are strongly-coupled and we have no reliable
  calculational methods. (QCD-like theories are already excluded.);

\item Extended technicolor theories must provide
a dynamicalexplanation of flavor.

\end{itemize}
\medskip\noindent
Ultimately, these problems are not likely to be solved without
{\bf experimental} direction.

\section*{Acknowledgments}
I thank Domenec Espriu and Toni Pich for organizing a stimulating summer
school.{\em This work was supported in part by the Department of Energy
  under grant DE-FG02-91ER40676.}

\end{document}